\documentclass[aps,pre,graphicx,nofootinbib,reprint,amssymb,showpacs,superscriptaddress,nobalancelastpage,tightenlines]{revtex4-2} 

\input{macros}

\usepackage[normalem]{ulem}

\graphicspath{{figs/}}

\begin{document}
\title{Surface fields and emergence of long-range couplings in classical uniaxial ferromagnetic arrays}
\author{D. B. Abraham}
\affiliation{Rudolf Peierls Centre for Theoretical Physics, Clarendon Laboratory, Oxford OX1 3PU, UK}

\author{A.~Macio\l ek}
\email{maciolek@is.mpg.de}
\affiliation{Institute of Physical Chemistry,
             Polish Academy of Sciences, Kasprzaka 44/52,
            PL-01-224 Warsaw, Poland}
\affiliation{Max-Planck-Institut f{\"u}r Intelligente Systeme, Heisenbergstr.~3,
D-70569 Stuttgart, Germany}

\author{A. Squarcini}
\affiliation{Institut f\"ur Theoretische Physik, Universit\"at Innsbruck,\\Technikerstra{\ss}e 21A, A-6020 Innsbruck, Austria }

\date{\today}
\begin{abstract}
Critical wetting is of crucial importance for the phase behaviour of a simple fluid or lsing magnet
confined between walls that exert opposing surface fields so that one wall favours liquid (spin up),
while the other favours gas (spin down). We show that arrays of boxes filled with fluid or Ising
magnet and linked by channels with such ``opposing'' walls can exhibit long-range cooperative effects,
on a length scale far exceeding the bulk correlation length. We give the theoretical foundations of
these long-range couplings by using a lattice gas (Ising model) description of a system.
\end{abstract}
\maketitle
\vfill

\section{Introduction}
\label{sec1}
Correlation  effects occurring over large distances are of great interest in  condensed matter physics from both fundamental and practical viewpoints. Gasparini and coworkers \cite{Perron_2010, Perron_2019} observed the emergence of long range couplings in experiments involving arrangements of bulk-like regions, i.e., boxes, of near superfluid ${}^4$He connected by small openings, such as  shallow channels.
Remarkably, even though the boxes had a mesoscopic spacing, various measurements  showed clear evidence of coupling between different boxes extending over distances much larger than the bulk correlation length.  
The authors of Ref.~\cite{Perron_2010}  made an interesting suggestion that ``action at a distance''~\cite{Fisher} effects of this type might be a more general feature of systems with phase transitions, both quantum (like ${}^4$He) and classical. 

\begin{figure}[htbp]
\centering
\includegraphics[width=0.3\textwidth]{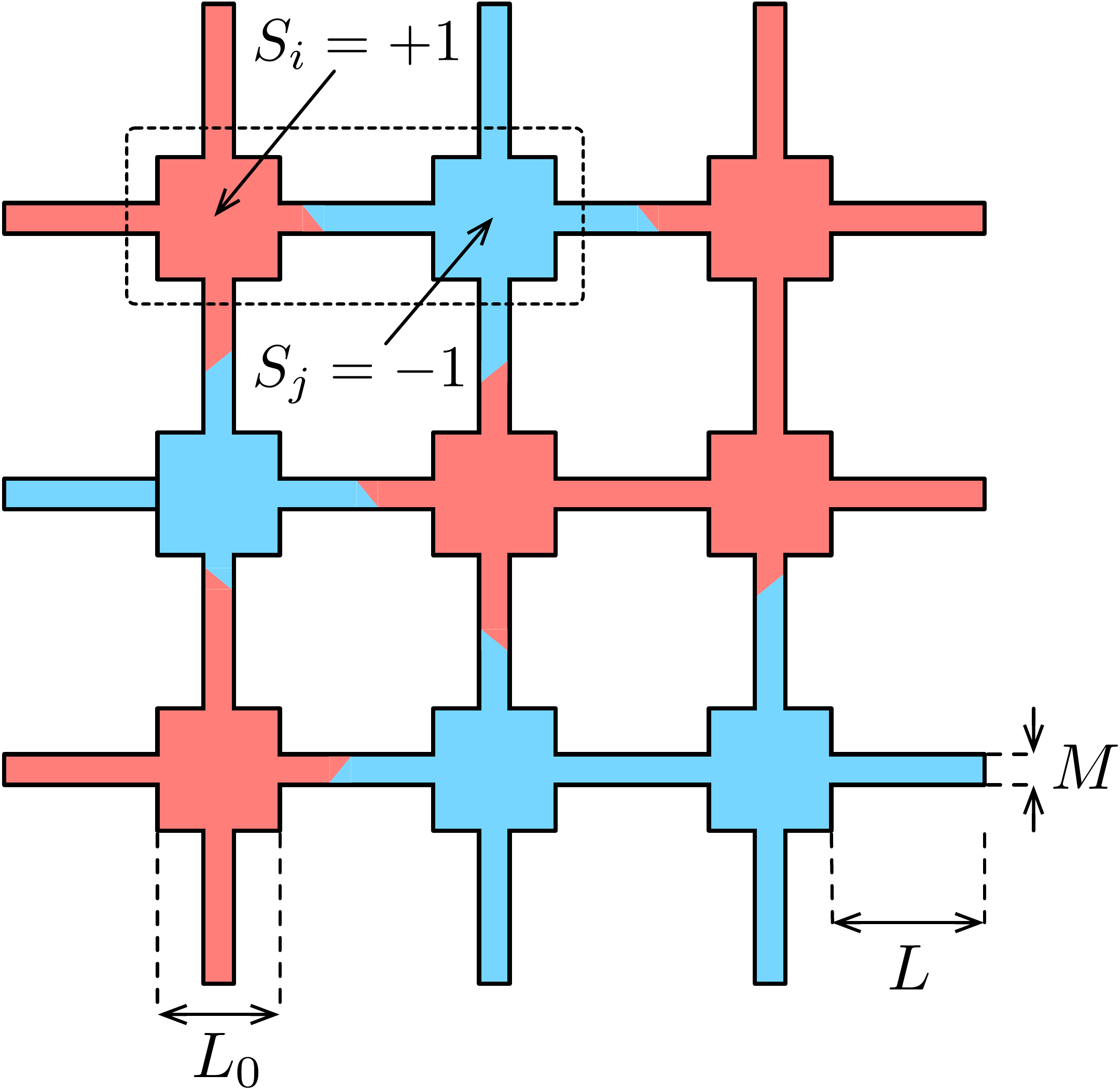}
\caption{Geometry of the two-dimensional array of cubes of size $L_0$ connected by channels (strips) of length $L$ and thickness $M\ll L$.  Different colors indicate  oppositely magnetized regions. In the ``network'' Ising model the state of a bulk-like box is describe by a spin variable $S_i=\pm 1$.}
\label{f01}
\end{figure}

Intrigued by these suggestions, we have recently presented a theoretical model of a classical system exhibiting correlation effects which are similar to  observations in superfluid helium~\cite{AMV_2014,AMSV_2017}. This model comprises of cubic Ising lattices of size $L_0^3$ arranged in a two-dimensional (2D) array and coupled together by Ising strips of size $L\times M, L\gg M$ (see Fig.~\ref{f01}). We have shown that, by appropriate tuning of temperature and size of the components of the array, the lattice of cubes develops long-range order, even though the connecting strips are very long compared to their lateral dimensions, just as in the case of Gasparini and coauthors. The arguments also work when the connecting strips are replaced by rods of length $L$ and cross section $M \times M$ with $L \gg M$. Physical realizations of our model include uniaxial classical ferromagnets or binary  mixtures in the lattice gas approximation (all belonging to the Ising model universality class of critical phenomena).

\begin{figure}[htbp]
\centering
\includegraphics[width=\columnwidth]{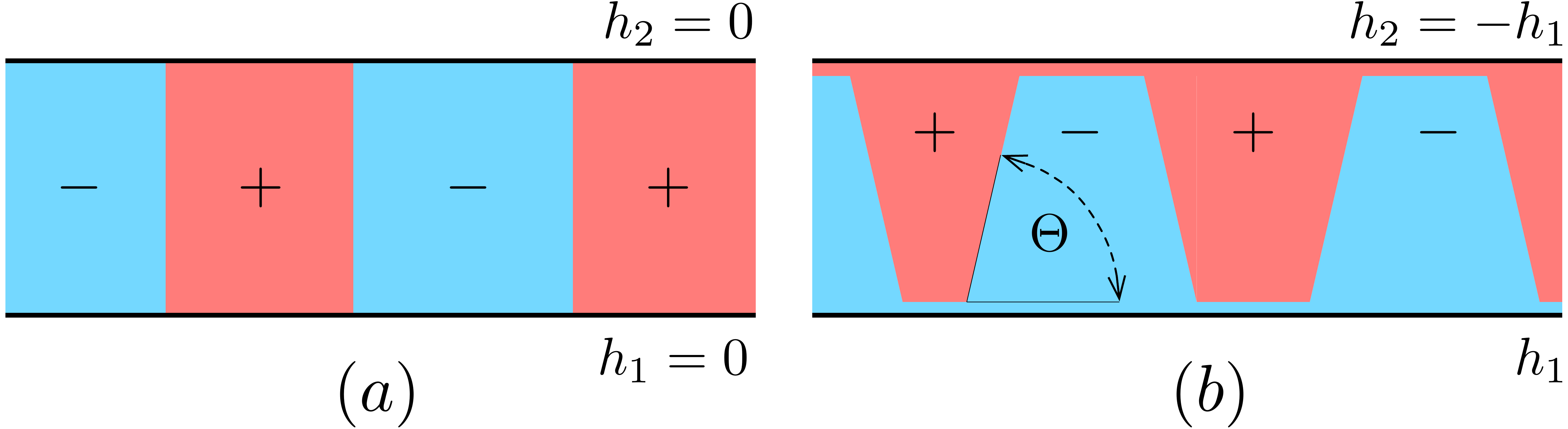}
\caption{Domain walls, which are Peierls contours after summing out fluctuations up to a scale of the bulk correlation length, in  a strip  (a) with free edges below the critical temperature $T_c$ and  (b) with opposing surface fields  at the edges below the wetting temperature $T_w(h_1)$.} 
\label{fig_strip}
\end{figure}

In the present paper we argue that the occurrence of action at a distance in Ising-like systems depends crucially on boundary conditions imposed on narrow ``connectors''  in a 2D array of large ``containers''. In  Refs~\cite{AMV_2014,AMSV_2017} we considered connecting channels with free boundary conditions. Below the critical temperature $T_c$, dominant spin configurations in such connectors involve Peierls contours~\cite{peierls1936ising,Griffiths} separating regions of alternating $(+)$ and $(-)$ magnetization stretching from side to side of the channel, as sketched in Fig.~\ref{fig_strip}(a).
The Peierls contours are responsible for breaking up long-range order in the channel on the characteristic length scale, which emerges from  asymptotic spectral degeneracy in transfer matrices~\cite{Kac}, and which is much larger than the bulk correlation length. The existence of such a characteristic length scale is intimately related to the appearance of the long-range order in 2D arrays. In the case of Ising strips or rods with all spins at the edges fixed, say at the $+1$ value,  there is no asymptotic spectral degeneracy in transfer matrices and hence  no long-range cooperative effects. In the presence of equal  and opposite surface fields, such degeneracy  appears in the partial wetting regime \cite{MS}. However, the mechanism of asymptotic degeneracy in this case is quite different from the mechanism in the strip with free edges.
In the partial wetting regime, the boundaries  are covered by either the $(+)$ or $(-)$ phase and  there is a single open Peierls contour separating the two oppositely magnetized regions touching the respective boundaries, as sketched in Fig.~\ref{fig_strip}(b). (It is clear that closed Peierls contour are allowed in the form of bubbles of phase $(-)$ in the cluster of phase $(+)$, and vice versa.) We can therefore expect long-range effects here, which  would open an intriguing possibility of tunability not only for uniaxial classical ferromagnets, but also for fluid systems and could be used, e.g., in as a microfabricated device in soft matter and biophysics experiments. 

Here, we explore this possibility by using a mesoscopic description, which is in the spirit of Fisher-Privman theory of finite-size effects at first-order  transitions~\cite{PF_1983} and which gives a simple physical picture of a typical spin configuration in a channel below the wetting temperature. 
By applying this description we reduce the problem to  a ``network'' planar Ising model, which focuses on the state of the cubes. For large enough cubes,  the state of each cube is characterized by $+1 $ and  $-1 $ magnetization. These ``boxes'' are coupled by channels in which the internal degrees of freedom have been summed out, producing an Ising superlattice of nodes with the effective coupling that are temperature dependent. In the case of 2D channels, we extend our mesoscopic description utilizing   exact results for the full Ising strip. This allows us to make a  quantitative prediction of the temperature range and the size of the connecting strips for which the ``network'' develops  long-range order.

The paper is organized as follows. Firstly, we show that in Ising strips with opposing surface fields $h_1=-h_2$, a characteristic divergent length scale develops below the wetting temperature $T_w(h_1)$. This can be inferred, e.g., from the decay of the spin-spin correlation function. In Sec.~\ref{sec2} we demonstrate that the spin-spin correlation function can be obtained from coarse-grained description  if one takes into account that in a partial wetting regime a typical spin configuration is one with domain walls inclined at the contact angle $\Theta$ to the edge of the strip (see Fig.~\ref{fig_strip}b)~\cite{{PhysRevLett.63.275}}. We derive the statistical weight of such an inclined domain wall in an independent exact calculation in the Ising strip (Appendix.~\ref{app:4}). In Sec.~\ref{sec3}  we construct the ``network'' planar Ising model and find the range of parameters for which it develops a long-range order. For comparison with the coarse-grained theory presented in Sec.~\ref{sec2}, in Sec.~\ref{sec4} we calculate the spin-spin correlation function  exactly using the transfer matrix method. In particular, we demonstrate how the ferromagnetic order is attained over distances of the order of characteristic length scale $\xi_{\parallel}$ that diverges exponentially with the cross section of the channels. The parallel correlation length scale $\xi_{\parallel}$ emerges as the inverse mass gap between two asymptotically degenerate modes in the spectrum of the transfer matrix with surface fields, that we obtain exactly. Our approach allows  us to treat the case of opposing surface fields ($h_{1}h_{2}<0$) in contrast to the technique used by Au-Yang and Fisher \cite{AuYangFisher_1980} that applies only to  case of  $h_{1}h_{2}>0$. Moreover, we construct surface states corresponding to two phases  pseudo-coexisting in the non-wet regime $T<T_w(h_1)$, as depicted in Fig.~\ref{fig_strip}. Note that for bulk 2D systems there can be no true phase transition  in the strip geometry. However, there is still a line of sharp (very weakly rounded) first-order phase transitions ending in the pseudocritical point~\cite{PF_1983,DEVIRGILIIS2005477}. Our conclusions are summarized in Sec.~\ref{sec5}.

\section{Strips with opposing surface fields: mesoscopic description}
\label{sec2}
\begin{figure}[htbp]
\centering
\includegraphics[width=8.7cm]{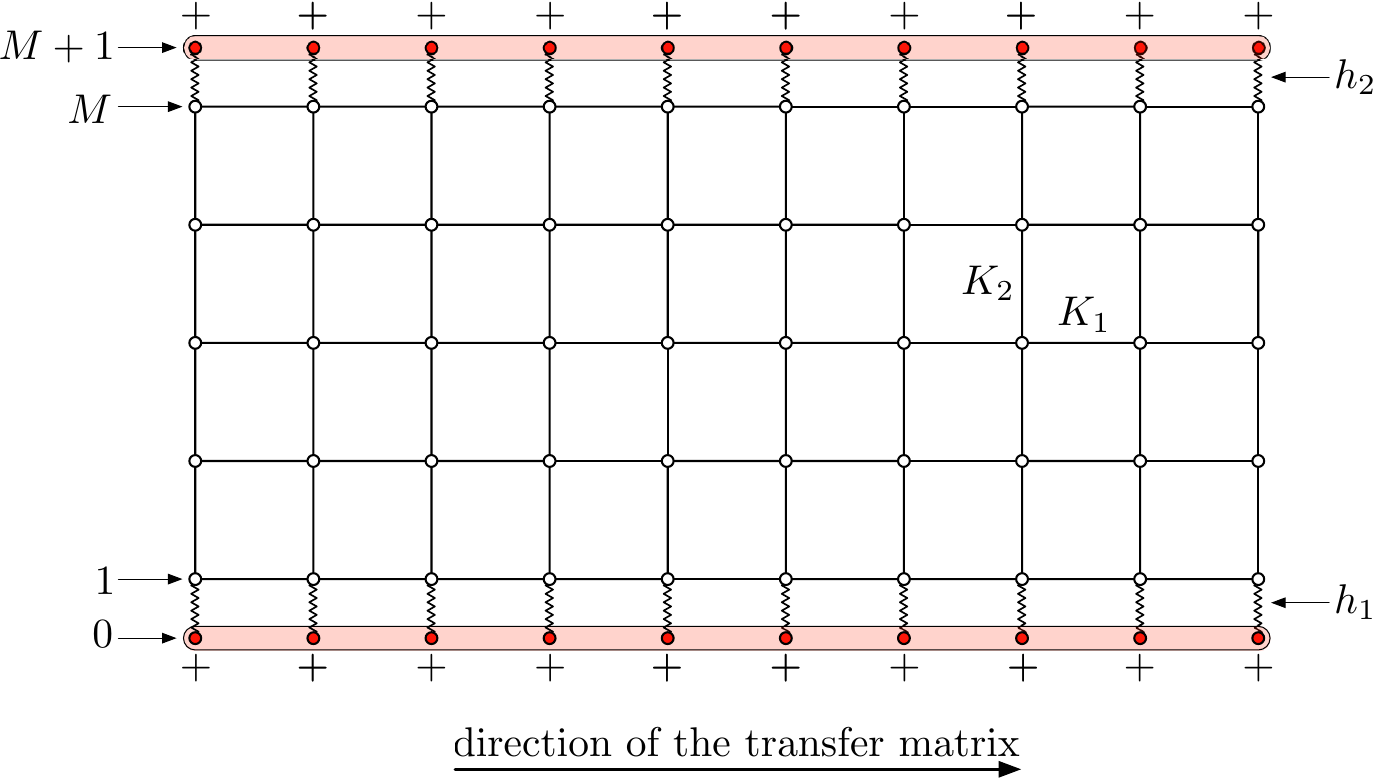}
\caption{Ising model on a rectangular lattice (strip)  with surface fields $h_{1}$ and $h_{2}$ as described in the text. The ghost rows with labels $m=0$ and $m=M+1$ are indicated in red.}
\label{f02}
\end{figure}

We start by considering the phase behavior of connecting channels with surface fields at the  boundaries. For a 2D Ising strip with surface fields $h_1=-h_2$ at the edges (shown in Fig.~\ref{f02}), at low temperatures one finds pseudo-coexistence at zero bulk field, where large domains of the ``bulk`` phases are formed~\cite{albano1989adsorption, Albano1990, PE_90, PE_92}. 
In this respect, the asymmetric strip behaves in the same fashion as the strip with free boundaries $h_1=-h_2=0$. At higher temperatures the influence of wetting manifests itself. For the semi-infinite square Ising model with a surface field $h_1$ Abraham's~\cite{Abraham_1980} exact solution shows that for a range of $h_1$ there is a critical wetting transition at a strictly subcritical temperature given by $w=1$ with
\begin{equation}
\label{e1}
w=\textrm{e}^{2K_{1}}( \cosh2K_{2} - \cosh2h_{1} )/\sinh2K_{2} \, ,
\end{equation}
and $0<h_1<K_2$. $K_{1}=\beta J_{1}$ and $K_{2}=\beta J_{2}$ are the coupling constants of interactions along vertical and horizontal bonds, respectively. In the region $w>1$, the interface is found on average at a finite distance from the wall, i.e.,  it is pinned. Above the transition, the interface depins to a fluctuating regime. In considering asymmetric strips it means that by choosing $h_1$ one can tune the temperature region of  pseudo-coexistence,  where large domains exhibiting thin ``wetting'' films exist. For example, if $h_1$ is weak pseudo-coexistence occurs almost all way up to the bulk critical temperature $T_c$. The same scenario should apply to the cubic Ising model in slab and rod geometry~\cite{PhysRevE.53.5023,PhysRevLett.90.136101,DEVIRGILIIS2005477,Kondrat}.

\begin{figure}[htbp]
\centering
\includegraphics[width=8.7cm]{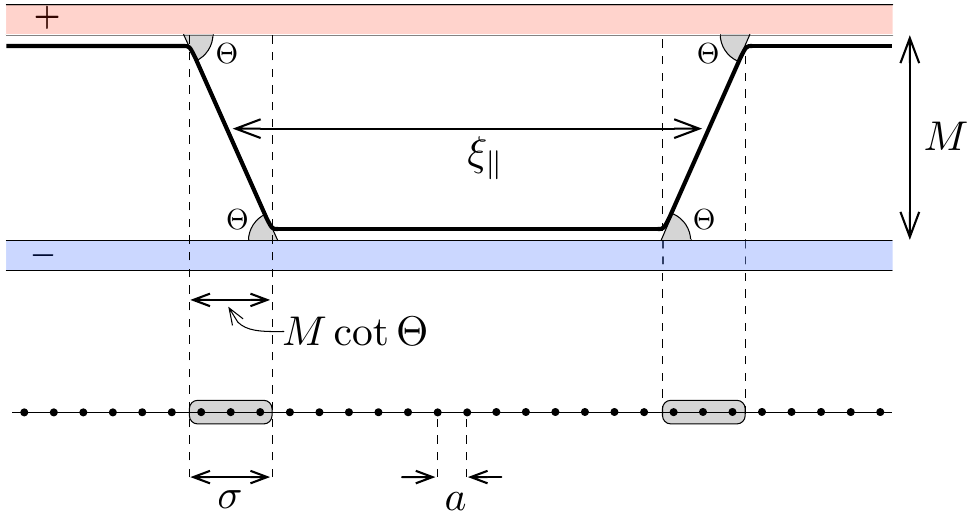}
\caption{A configuration of domain walls in the channel with wetting boundaries.
Domain walls form a contact angle $\Theta$ with the solid wall and $\xi_{\parallel}$ provides a measure of their separation. The effective picture is that of one-dimensional lattice gas of particles with diameter $\sigma=1+[M\cot\Theta]$. }
\label{f03}
\end{figure}

\begin{figure}[htbp]
\centering
\includegraphics[width=8.3cm]{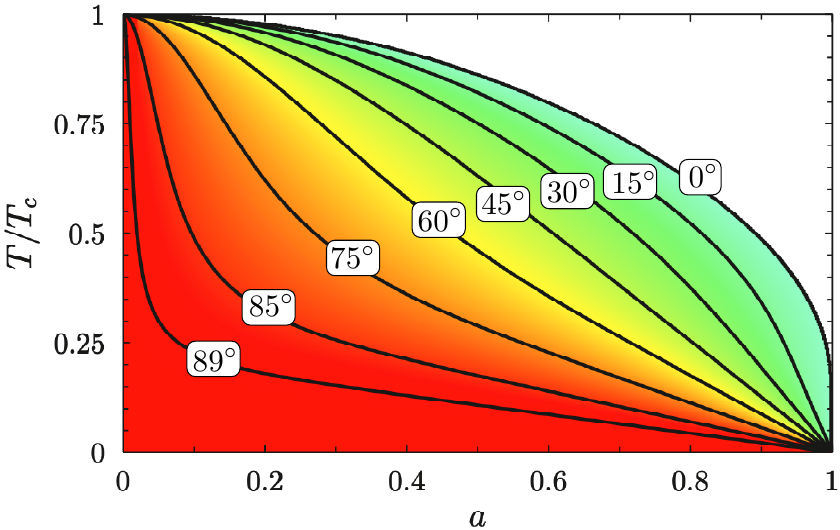}
\caption{The contact angle $\Theta$ as function of the rescaled field $a=\vert h_{1}\vert /K_{2}$ and temperature $T$. The locus with $\Theta=0$ corresponds to the wetted phase boundary where $w(h_{1},T)=1$. In this figure $K_{1}=K_{2}$. The contact angle tends to $\pi/2$ for any subcritical temperature provided the surface field tends to zero, therefore for free boundaries $\Theta(h_{1}=0,T)=\pi/2$, and domain walls are perpendicular to the strip edges.}
\label{f04}
\end{figure}

On the scale of bulk correlation length, a typical configuration at pseudo-coexistence is one with regions of alternating $(+)$ and $(-)$ magnetization, with a magnitude roughly the spontaneous magnetization, separated by domain walls. Unlike a strip with free boundaries, where  domain walls run perpendicular to the edges to minimize an energetic penalty proportional to their length, in a strip with surface fields they are inclined  at the contact angle $\Theta(h_1,T)$ as shown in  Fig.~\ref{f03}. 
The exact expression for the contact angle is~\cite{PhysRevLett.63.275}
\begin{equation}
\label{e2}
\tan\Theta = \frac{ w^{2}-1 }{ [(A-w)(B-w)(w-A^{-1})(w-B^{-1})]^{1/2} } \, .
\end{equation}
Figure~\ref{f04} shows $\Theta$ as function of the variables $h_{1}$ and $T$. The derivation of $\Theta(h_{1},T)$ is carried out in Appendix~\ref{app:1}, there the meaning of the notation for $A$ and $B$ is explained.

As first pointed out by Parry and Evans \cite{PE_90,PE_92}, the characteristic length $\xi_{\parallel}$ of successive domains of $(+)$ and $(-)$ magnetization in the non-wet regime should diverge exponentially with the width of the strip. This prediction is consistent with exact results from  the restricted solid-on-solid model of an interface, where $\xi_{\parallel}$ is obtained from the two largest eigenvalues of the transfer matrix~\cite{PhysRevB.37.3713,Selke}. In Sec. \ref{sub:lengthscale} we show that for a 2D Ising strip the exact result for $\xi_{\parallel}$ is
\begin{equation}
\label{ }
\xi_{\parallel} = \frac{(Aw-1)(Bw-1)}{2w\sqrt{AB}\left(w-w^{-1}\right)^{2}} w^{M} \, ;
\end{equation}
the exponential growth follows from $w>1$. We  determine this length  from the decay of the correlation function $\mathcal{G}_M(m,n)$ of two spins in the $m$th row separated by $n$ columns (see Fig.~\ref{f02}). We calculate $\mathcal{G}_M(m,n)$ using the transfer matrix method. In the limit of $L\to \infty$  and for finite $M$, the leading asymptotic decay of $\mathcal{G}_M(m,n)$ for  $n \gg M$ is  given by $\left(\Lambda_{2}/\Lambda_{1}\right)^n = \exp\left(-n/\xi_{\parallel}\right)$ (see, e.g. \cite{thompson2015mathematical}), where $\Lambda_{1}$ and $\Lambda_{2}$ are the two largest eigenvalues of the transfer matrix. The analysis of the spectrum of the transfer matrix presented in Sec.~\ref{subsec:AD} shows that  below the wetting temperature these two eigenvalues correspond to the imaginary wavenumbers $k_1=iv_1$ and $k_2=iv_2$, which are  asymptotically degenerate. Thus  the parallel correlation length is determined by  the inverse mass gap between two asymptotically degenerate imaginary modes
\begin{equation}
\label{e36}
\xi_{\parallel}^{-1} = \gamma(iv_{2})-\gamma(iv_{1}) \, .
\end{equation}
The Onsager function~\cite{Onsager_44} $ \gamma(iv_{k}), k=1,2$ is calculated from $\cosh \gamma(iv_k) = \cosh 2K_1^*\cosh 2K_2 - \sinh 2K_1^*\sinh 2K_2\cosh v_k $, where $\exp(-2K_k^*)= \tanh K_k$. The microscopic analysis of the two creation operators for the two imaginary wavenumber modes  (see Sec.~\ref{sub:surfst} for details),  enables us to write down states which locate the interface  near either one edge or near the other. Thus these states, which are  "not quite" eigenvectors of the transfer matrix in the diagonalisation, correspond to two phases pseudo-coexisting in the partial wetting  regime. 

The exponential divergence of  $\xi_{\parallel}$ can also be inferred in a simple  way by using a coarse-grained description based on domain walls. To this end, we treat the collection of domain walls for the strip geometry as a quasi-one-dimensional gas of strictly avoiding particles on a lattice. To account for the slope of the domain wall, we assume that the particles are not point-like, but have a diameter $\sigma$. The domain wall  projection onto the edge of the strip  is equal to  $M\cot\Theta$ as shown in Fig.~\ref{f03}. On a lattice with the lattice constant $a=1$ we set $\sigma= [M\cot\Theta]+1$, where the symbol $[x]$ denotes the integer part of $x$, i.e., $x$ is the greatest integer less than or equal to $x$. If the surface fields vanish, then $\sigma = a$ so the domain wall  becomes a pointlike particle, which is the case we studied in Refs.~\cite{AMV_2014,AMSV_2017}.

The equilibrium statistical mechanics of this system can be determined within the grand canonical ensemble. If $\zeta$ is the fugacity corresponding to the Boltzmann weight associated with an isolated domain wall, then the grand partition function for a strip of length $L$ is
\begin{equation}
\label{e3}
\Xi^{\rm rods}(\zeta,L,\sigma) = \sum_{j=0}^{[L/\sigma]} \binom{L-j\sigma+j}{j} \zeta^{j} \, ,
\end{equation}
where $[L/\sigma]$ is the maximum number of particles that can be allocated on the strip of length $L$. The binomial coefficient counts all possible arrangements of $j$ indistinguishable hard rods of length $\sigma$ on a lattice of length $L$. The rods are hard in the sense that they can touch but no overlap. For free boundary conditions the diameter  becomes a lattice unit, then the combinatorial factor in Eq.~(\ref{e3}) reduces to the binomial $\binom{L}{j}$ \cite{AKS_1989}.

Since the distance between the domain walls is large, we can approximate the grand partition function $\Xi^{\rm rods}$ by reducing the problem of hard rods on a lattice to the one of point particles on a coarse lattice with a lattice constant of $\sigma$. In fact, the problem is reduced to the study of the diluted limit of a hard rod gas. (For a formal details of this approximation as well as the connection to the continuum version of the model known as  the Tonks-Rayleigh hard rod gas~\cite{Tonks,RobledoRowlinson}, see Appendix~\ref{app:2}.)
This approximation gives
 \begin{equation}
\label{e4}
\Xi(\zeta,L,\sigma) \approx \sum_{j=0}^{\ell}\binom{\ell}{j} \sigma^{j} \zeta^{j} = (1+\widetilde{\zeta})^{\ell}\, ,
\end{equation}
where $\widetilde{\zeta}=\sigma\zeta$ and  $\ell=L/\sigma$.

Within this simple physical picture, the calculation of the pair correlation function $G(x)$ for a separation $\sigma x$ is straightforward. If a pair of spins is  parallel (antiparallel), they are separated by an even (odd) number of domain walls. Denoting by $\Xi_e(x)$ and $\Xi_o(x)$ the grand partition function with an even and odd number of particles, respectively, we find
\begin{equation}
\label{e5}
G(x) \propto \frac{\Xi_e(x)-\Xi_o(x)}{\Xi_e(x)+\Xi_o(x)}= \frac{(1+\widetilde{\zeta})^{x}}{(1-\widetilde{\zeta})^{x}} \, .
\end{equation}
The proportionality factor is not obtainable within this approach since on the mesoscopic scale, the local magnetization at the correlated sites is not $\pm 1$. For  $\widetilde{\zeta} \ll 1$, the expression for $G(x)$ can be simplified giving a purely exponential decay
\begin{equation}
\label{e6}
G(x) \sim G_0 \exp \left[-2x\widetilde{\zeta}\left(1+O(\widetilde{\zeta}^2)\right)\right] \, .
\end{equation}
This is consistent with  Landau's argument about the lack of long-range order in a one-dimensional system with short-range interactions. 
The statistical weight of a domain wall  is taken \emph{a priori} $\zeta = \exp(-\mathcal{F}(\Theta))$, where $\mathcal{F}(\Theta)$ is the energy cost  associated with the insertion of a domain wall into the strip with opposing surface fields. This energy cost is proportional to $M$, hence, the mesoscopic description predicts that the decay length of the correlation function $G(x)$ \textit{diverges exponentially with $M$}. In Appendix~\ref{app:1} we argue that $\mathcal{F}(\Theta)=Mv_0$ with $Mv_{0}  =   M\csc\Theta \, \tau(\Theta)-M\cot\Theta f_{0}$, where $\tau(\Theta)$ is the angle-dependent surface tension~\cite{AR74,AU88} for an inclined interface forming a tilt angle $\Theta$ with the wall and $f_{0}$ is the surface free energy of a flat interface pinned  to the wall.

\begin{figure}[htbp]
\centering
\includegraphics[width=7.9cm]{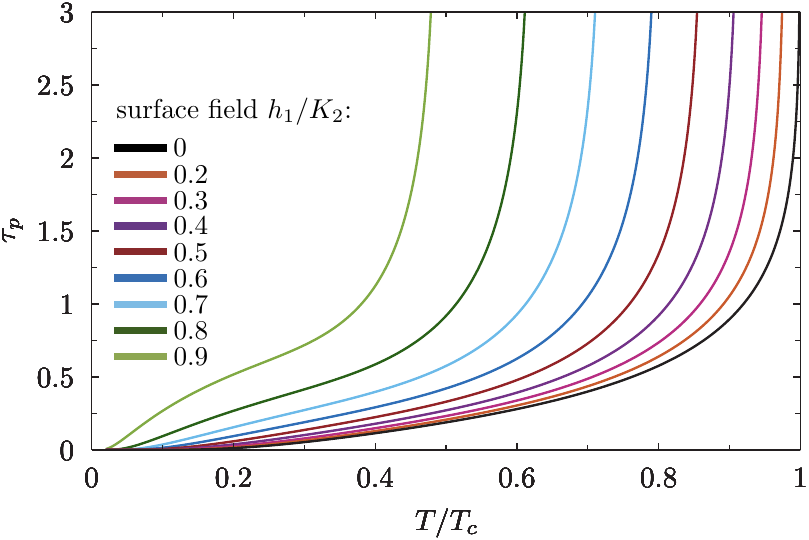}
\caption{The point tension $\tau_{p}$ as function of temperature $T$ for various surface fields  (as shown in the inset). The rightmost curve (black) corresponds to the  free edges $a=0$. The point tension diverges logarithmically at the wetting temperature $T_w(h_1)$. In this figure, $K_{1}=K_{2}$.
}
\label{f05}
\end{figure}

We now compare this prediction with the results of exact calculations for the full Ising strip, which we present in Sec.~\ref{sec4}. We find agreement in the asymptotic behavior of $G(x)$ provided $\zeta=\exp(-Mv_{0})$ is replaced by
\begin{equation}
\label{e7}
\zeta (T,h_1,M) = \frac{w\sqrt{AB}\left(w-w^{-1}\right)^{2}}{(Aw-1)(Bw-1)} \textrm{e}^{-Mv_0} \, .
\end{equation}
The prefactor in Eq.~(\ref{e7}) is due to the point tension $\tau_{p}$ (a 2D analogue of the line tension)~\cite{Abraham_1993} arising at the points where flat portions of the domain wall meet  the inclined one. We calculate this prefactor exactly in Appendix~\ref{app:4}. The point tension $\tau_{p}$ as a function of temperature for different values of the surface field is shown in Fig.~\ref{f05}.

Finally, exact calculations presented in Sec.~\ref{sec4} allow for the following identification
\begin{equation}
\label{e8}
\zeta(T,h_1,M) = \textrm{e}^{-2\tau_{p}}\textrm{e}^{-Mv_0}= \xi_{\parallel}^{-1}/2 \, ,
\end{equation}
where $\xi_{\parallel}$ is given by Eq.~(\ref{e46}). Moreover, if we consider spins at distance $m$ from the channel and compare Eq.~(\ref{e6}) with the exact calculation for the full Ising strip, then the prefactor $G_0$ of the exponential decay can be identified with the amplitude 
$\bar{\mathfrak{m}}_{m}^{2}(M)$ in Eq.~(\ref{e45}). 
\section{Network planar Ising model}
\label{sec3}
We now apply the mesoscopic description to a pair of cubic lattice boxes of side $L_0$ coupled by an Ising strip of dimension $L \times M$, with $L_0 \gg M$. For $T<T_w$, the picture which emerges is one with  a sequence of inclined domain walls crossing the strip,
but none inside the boxes. The domain walls  intersecting the boxes are of size  $\sim L_0^{d-1}$ and are therefore  suppressed due to the much higher cost of free energy. Because  boxes  are large, we expect that below the wetting temperature $T_w(d=2)$, which is lower than the critical temperature $T_c(d=2)$, which in turn is lower than $T_c(d=3)$, the state of each box is either magnetized up or down. Within our coarse-grained description, we can assign a variable $S_j =\pm 1$ for each box as illustrated in Fig.~\ref{f01}, and calculate a strip-mediated
effective interaction energy  $K_{\textrm{eff}}S_iS_j$ for a given argument of the $S_j$. This can be done by noticing that if  spins on neighboring boxes $i$ and $j$ are parallel, that is $S_iS_j = 1$, then there must be an even number of domain walls on the connecting strip which has length $L$. On the other hand, if $S_iS_j = -1$, the location of the spins is separated by an odd number of domain walls. As follows from the previous  section, the grand partition function with an even number of particles, denoted $\Xi_e(\ell)$ for a lattice of length $L=\sigma \ell$, is just
 \begin{equation}
 \label{e9}
 \Xi_e(\ell) = \sum_{m=0, \rm{even}}^{\ell}\binom{\ell}{m}\widetilde{\zeta}^{m}= 2^{-1}\left\{(1+\widetilde \zeta)^{\ell} + (1-\widetilde \zeta)^{\ell}\right\} \, .
\end{equation}
The analogous result for an odd number of particles is
 \begin{equation}
 \label{e10}
 \Xi_o(\ell) = \sum_{m=0, \rm{odd}}^{\ell}\binom{\ell}{m}\widetilde{\zeta}^{m}= 2^{-1}\left\{(1+\widetilde \zeta)^{\ell} - (1-\widetilde \zeta)^{\ell}\right\} \, .
\end{equation}
Thus the  weight of a strip for given spin variables $S_i, S_j$ can be written in an Ising form
\begin{equation}
\label{e11}
B(S_i,S_j) = \Xi_e^{(1+S_iS_j)/2}\Xi_o^{(1-S_iS_j)/2}=\mathcal{A}e^{K_{\textrm{eff}}S_iS_j} \, ,
\end{equation}
where $\mathcal{A}^2 = \Xi_e(\ell)\Xi_o(\ell)$ and the coupling $K_{\textrm{eff}}$ is given by $e^{2K_{\textrm{eff}}} = \Xi_e(\ell)/\Xi_o(\ell)$. The equation for the coupling $K_{\textrm{eff}}$ can be re-written as:
\begin{equation}
\label{e12}
\ell \ln\left[(1+\widetilde{\zeta})/(1- \widetilde{\zeta})\right] = \ln\coth K_{\textrm{eff}} \, .
\end{equation}
We  stress here that the temperature evidently does not enter in the usual Boltzmann way in $K_{\textrm{eff}}$, which has interesting physical consequences, as will be shown in the following.

\begin{figure}[htbp]
\centering
\includegraphics[width=0.4\textwidth]{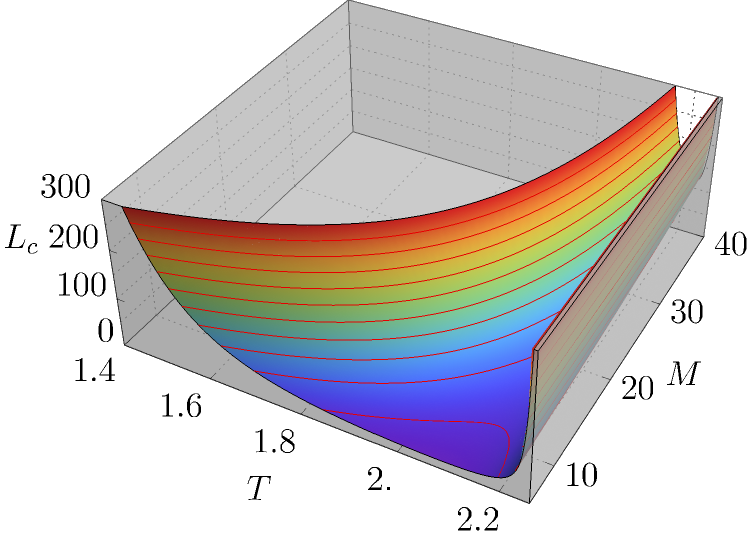}
\includegraphics[width=0.4\textwidth]{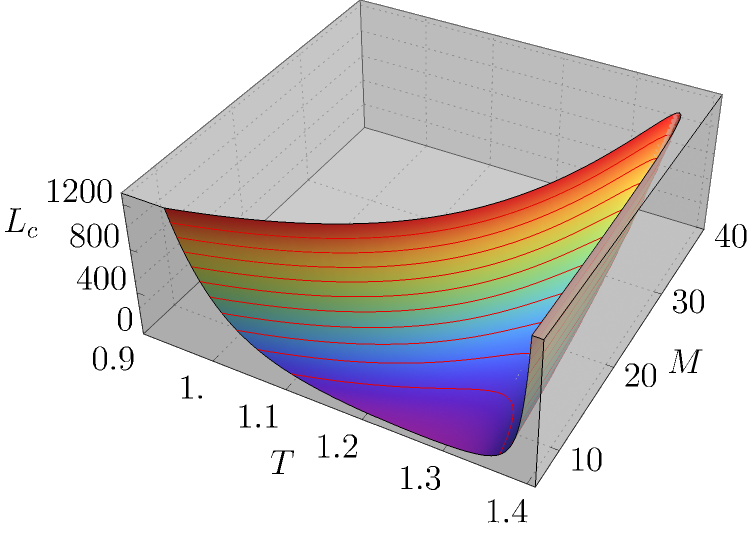}
\caption{The critical value of the  length of connecting strips $L_{c}$ of the two-
dimensional ``network'' lattice shown in Fig.~\ref{f01} as a function of the width of the strip $M$ and temperature $T$ for two values of the reduced surface field (a) $h_{1}/K_{2}=0.1$ corresponding to $T_w=2.2571$ and (b) $h_{1}/K_{2}=0.8$ corresponding to $T_w=1.40966$. The network is ordered in the grey region lying below the critical surface $L_c(M,T,h_1)$.} 
\label{f06}
\end{figure}

Having constructed effective bonds, we can now assemble them and boxes to make up a two-dimensional ``network`` lattice as illustrated in Fig.~\ref{f01}. The long-range order will appear in this network, if the parameters can be tuned such that $K_{\textrm{eff}}$ satisfies $K_{\textrm{eff}} > K_{c}(d=2)=(1/2)\ln(1+\sqrt{2}) \approx 0.440687$~\cite{Onsager_44}. Thus, if the geometrical parameters $L$, $M$, and the intensive thermodynamic variables $h_1$ and $T$ satisfy the following inequality
\begin{equation}
\label{e13}
 \ell\ln\left[(1+\widetilde{\zeta})/(1- \widetilde{\zeta})\right] < \ln\left(1+\sqrt 2\right) \, ,
\end{equation}
then the network lattice is ferromagnetically ordered. Given any integer-valued $M$ of the width of the connecting strip, the surface field $h_1$ and temperature $T< T_w(h_1)$, it is always possible to choose a critical integer-valued $\ell_c$ for which left and right hand side of Eq.~(\ref{e13}) are equal. Because $\widetilde{\zeta}$ is small away from the wetting temperature ($w=1$), we can write
\begin{equation}
\label{e14}
\frac{ L_c}{\xi_{\parallel}} = \ln\left(1+\sqrt 2\right) \, ,
\end{equation}
which implies that $L_c$ diverges exponentially with $M$. The phase diagram of the two-dimensional ``network'' lattice shown in Fig.~\ref{f06} for two values of the surface fields corresponding to $T_w=2.2571$ ($\vert h_{1}\vert /K_{2}=0.1$) and $T_w=1.40966$ ($\vert h_{1}\vert /K_{2}=0.8$). Notice that for temperatures $T/T_w \lesssim 0.8$, the bulk correlation length does not exceeds $\xi_b \approx 2$, while already at $M = 12$ and $T /T_w$ = 0.8 the critical length is $L_c \approx 400 $ for $\vert h_{1}\vert /K_{2}=0.1$ and  $L_c \approx 4050$ for $\vert h_{1}\vert /K_{2}=0.8$.

 Equation (\ref{e14}), which determines the phase boundary between order and disorder phase of array of boxes 
in a parameter space spanned by temperature and size of connecting strip is the central result of this paper. 

\section{Strips with opposing surface fields: Exact results}
\label{sec4}
In this section we determined the length $\xi_{\parallel}$ exactly using the transfer matrix method. Moreover, we define the surface states, which produce the asymptotic degeneracy of the transfer matrix, and hence the divergence of $\xi_{\parallel}$. We now specify the model more precisely.

\subsection{Ising model and Transfer Matrix}
\label{sec:model}
We consider a planar Ising ferromagnet  in strip geometry with a zero  magnetic field. We introduce lines of weakened bonds $|h_1|, |h_2| < K_2$ normal to and contiguous with the surfaces  as shown in Fig.~\ref{f02}. We set $\sigma_{n,0}=\sigma_{n,M+1}=+1$ and allow  $h_{1}$ and $h_{2}$ to take both signs. In the following, we only consider the perfectly antisymmetric cases $h_1=-h_2$. We construct transfer matrix working in $(1,0)$ direction, i.e., parallel to the edges. The column  to column (see Fig.~\ref{f02}) transfer operator is given by:
\begin{equation}
\label{e15}
\textsf{V}_{1} = (2\sinh 2K_{1})^{M/2} \exp \biggl[ - K_{1}^{\star} \sum_{m=1}^{M}\sigma_{m}^{z} \biggr] \, ,
\end{equation}
where $\tanh K_{1}=\textrm{e}^{-2K_{1}^{\star}}$. The intra-row couplings along the surface rows $m=0$ and $m=M+1$ is $K_{0}\rightarrow\infty$. The dual coupling $K_{0}^{\star}\rightarrow0$, therefore the indexes $m=0$ and $m=M+1$ do not report in (\ref{e15}). Here, $\sigma_m^i, i=x, y, z$ are spin operators\footnote{
The operator $\sigma_{m}^{\alpha}$ acts on the tensor product of the two-dimensional Hilbert spaces of each lattice site in a column
\begin{equation}
\label{e16}
\sigma_{m}^{\alpha} = \Mybigotimes_{j=1}^{m-1} \mathbf{1} \Mybigotimes \sigma^{\alpha} \Mybigotimes_{j=m+1}^{M} \mathbf{1} \, ,
\end{equation}
where $\sigma^{\alpha}, \alpha=x,y,z$ are Pauli matrices. In particular $\sigma_{m}^{\alpha}$ fulfills the on-site anti-commutation relation
\begin{equation}
\label{e17}
[\sigma_{m}^{\alpha}, \sigma_{m}^{\beta}]_{+} = 2\delta_{\alpha\beta} \, ,
\end{equation}
while for $n \neq m$, $[\sigma_{m}^{\alpha}, \sigma_{n}^{\beta}]_{-}=0$.} with {\it ordered} direction taken as $x$, i.e., the magnetization is given by the average of $\sigma_m^x$~\cite{SML}. The transfer matrix $\textsf{V}_{2}$ which accounts for the interactions within columns is of the diagonal form
\begin{equation}
\label{e18}
\textsf{V}_{2} = \exp \biggl[ h_{1}\sigma_{0}^{x}\sigma_{1}^{x} + K_{2} \sum_{m=1}^{M-1}\sigma_{m}^{x}\sigma_{m+1}^{x} + h_{2}\sigma_{M}^{x}\sigma_{M+1}^{x} \biggr] \, .
\end{equation}  
As noted by Kaufman \cite{Kaufmann_49}, in order to diagonalise the symmetrised forms of the transfer operators it is essential to introduce the Jordan-Wigner transformation and the lattice spinors $\Gamma_{m}$ defined by
\begin{equation}
\begin{aligned}
\label{e19}
\Gamma_{2m-1} & = \mathcal{P}_{m-1} \sigma_{m}^{x} \, , \\
\Gamma_{2m} & = \mathcal{P}_{m-1} \sigma_{m}^{y} \, , \qquad {\rm for} \, \, m=1,\dots, M+1 \, ,
\end{aligned}
\end{equation}
supplemented with $\Gamma_{-1} = \sigma_{0}^{x}$ and $\Gamma_{0} = \sigma_{0}^{y}$, and where
\begin{equation}
\label{e20}
\mathcal{P}_{m} = \prod_{j=0}^{m}\left( -\sigma_{j}^{z} \right) \, ,
\end{equation}
is the so called ``fermionic tail''. The spinors we just introduced are self-adjoint operators with square equal to unity, they anticommute with each other and fulfill the Clifford algebra.
\begin{equation}
\label{e21}
[\Gamma_{m} , \Gamma_{n}]_{+} = 2\delta_{m,n} \, .
\end{equation}
The spinors are related to the fermionic operators $X(k), k=1, \ldots, M$ by
\begin{equation}
\label{e22}
X(k) = \frac{1}{2} N(k) \sum_{m=0}^{2M+1}y_{m}(k)\Gamma_{m} \, ,
\end{equation}
where the normalization factor $N(k)$  ensures the canonical anti-commutation relation, i.e., $[X(k_{1}),X^{\dag}(k_{2})]_{+}=\delta_{k_{1,k_{2}}}$. The functions $y_{m}(k)$ have to be determined such that in terms of $X(k)$ and $X^{\dag}(k)$ the transfer operator $\textsf{V}=\textsf{V}_2^{1/2}\textsf{V}_1\textsf{V}_2^{1/2}$ admits the diagonal form 
\begin{equation}
\label{e23}
\textsf{V} = \exp\biggl[ - \sum_{k \in \Omega_{M} } \gamma(k) \left( X^{\dag}(k)X(k) -1/2 \right) \biggr] \, .
\end{equation}
The Onsager function \cite{Onsager_44} $\gamma(k)$ is the non-negative solution of
\begin{equation}
\label{e24}
\cosh \gamma(k) = \cosh 2K_1^*\cosh 2K_2 - \sinh 2K_1^*\sinh 2K_2\cos k \, ,
\end{equation}
 for real $k$. Note that operators associated with $\Gamma_{-1}$ and $\Gamma_{2(M+1)}$, $X_0=(1/2)\left(\Gamma_{-1} + i\Gamma_{2(M+1)}\right)$ and its conjugate, do not appear in (\ref{e23}). They are zero-energy operators, which  implies that each eigenvalue is doubly degenerate. We denote the vacuum\footnote{ The vacuum  $\vert \Phi_{\infty}\, \rangle = X_0 \vert \Phi \, \rangle$, where  $\vert \Phi \, \rangle $  is a vacuum determined by $X(k)\vert \Phi \, \rangle = 0$ for all $k \in \Omega_{M}$.} for the operators $X(k)$ and $X_0$, which is also the maximal eigenvector with eigenvalue $\Lambda_0$, by $\vert \Phi_{\infty} \, \rangle$. The vacuum $\vert \Phi_{\infty} \, \rangle$ includes  all four possible cases in which the spins in a given edge are parallel. The spectrum for the edge state corresponding  to $h_1<0$ and $h_2>0$ is constructed with the use of appropriate projectors ~\cite{AR74,MS}
\begin{equation}
\label{e25}
X^{\dag}((k)_{2n+1}) \vert ++ \rangle \, ,
\end{equation}
where the notation in the above is:
\begin{equation}
\label{e26}
X^{\dag}((k)_n) = X^{\dag}(k_1)\cdots X^{\dag}(k_n) \, ,
\end{equation}
and
\begin{equation}
\label{e27}
\vert ++\rangle = 2^{-1/2}(1+X^{\dag}_0) \vert \Phi_{\infty} \, \rangle \, ,
\end{equation}
is the normalized  state with plus boundary spins. For the case of  $h_1>0$ and $h_2<0$, the spectrum is constructed by replacing $\vert ++\rangle$ in Eq.~(\ref{e27})
by the state $\vert -- \rangle = 2^{-1/2}(1 - X^{\dag}_0) \vert \Phi_{\infty}\, \rangle$. 
The sum appearing in (\ref{e23})  is restricted to wavenumbers $k$ compatible with the boundary conditions on the edges of the strip. 
This generates the following discretization condition \cite{MS}
\begin{equation}
\label{e28}
\textrm{e}^{iMk} = s\textrm{e}^{i\delta(k)}=s\textrm{e}^{i\delta^{\prime}(k)}\frac{w\textrm{e}^{ik}-1}{\textrm{e}^{ik}-w}   \, ,
\end{equation}
whose solutions  define the set $\Omega_{M}$. The parity number $s=\pm 1$ encodes reflection behavior of the eigenvectors, $\delta^{\prime}(k)$ is the angle introduced by Onsager (see Appendix~\ref{app:3}) and $w$  is the wetting parameter [Eq.~(\ref{e1})].   For the $\textsf{V}$ symmetrization we have 
\begin{equation}
\begin{aligned}
\label{e29}
y_{2m+1}(k) = & - \textrm{e}^{-i\delta^{*}} \textrm{e}^{imk} + \textrm{e}^{i\delta} \textrm{e}^{-imk}  \, , \\
y_{2m}(k) = & i \left( - \textrm{e}^{imk} + \textrm{e}^{i\delta}\textrm{e}^{-i\delta^{*}} \textrm{e}^{-imk} \right) \, , \\
m = & 1,\ldots, M-1\, ,
\end{aligned}
\end{equation}
with the boundary values
\begin{equation}
\begin{aligned}
\label{e30}
y_{0}(k) = & i\sqrt{\frac{B}{B-w}}\left(-1 + \textrm{e}^{i\delta}\textrm{e}^{-i\delta^{*}}\right) \, , \\
y_{1}(k) = & \sqrt{\frac{A}{A-w}}\left(-\textrm{e}^{-i\delta^{*}} + \textrm{e}^{i\delta}\right) \, ,
\end{aligned}
\end{equation}
where $A=\exp[2(K_1+K_{2}^{\star})]$ and $B=\exp[2(K_1-K_{2}^{\star})]$. The quantities $y_{2M}(k)$ and $y_{2M+1}(k)$ are obtained by using the reflection symmetry:
\begin{equation}
\begin{aligned}
\label{e31}
& y_{2(M+1-m)} =  -is \, y_{2m-1} \, , \\
& y_{2(M-m)+1} =  i s \, y_{2m} \, .
\end{aligned}
\end{equation}

\subsection{Asymptotic degeneracy}
\label{subsec:AD}
In order to proceed it is crucial to discuss the allowed momenta $k$ for subcritical temperatures $K_2 > K_1^{\star}$. We restrict $k$ to the range $[0,\pi]$.  At $T_w(h_1)$ there is a special solution at $k=0$ with nonzero eigenvector and the corresponding eigenvalue $\Lambda_0 \exp[\gamma(0)]$. For all other temperatures the values $k=0$ and $k=\pi$ give trivial eigenvectors. In the partially wet regime $T< T_w(h_1)$ there are $M-2$ real solutions between $0$ and $\pi$. Two solution of the discretization condition (\ref{e28}) are found at imaginary values $k_1=iv_1$ and $k_2=iv_2$~\cite{MS}:
\begin{equation}
\begin{aligned}
\label{e32}
v_{1} & \simeq v_{0} - \mathcal{A}(T,h_{1}) w^{-M} \, , \qquad (s=+1) \\
v_{2} &  \simeq v_{0} + \mathcal{A}(T,h_{1}) w^{-M} \, , \qquad (s=-1) \, ,
\end{aligned}
\end{equation}
with $v_{0}=\ln w$ and
\begin{equation}
\label{e33}
\mathcal{A}(T,h_{1}) = \left(\frac{A-w}{Aw-1} \frac{B-w}{Bw-1}\right)^{1/2} \left(w-w^{-1}\right) \, .
\end{equation}
The symbol $\simeq$ in Eq.~(\ref{e32}) stands for the omission of subdominant terms of order $w^{-2M}$. Note that below bulk criticality $(B>1)$ and within the wetting regime ($w>1$) each factor in (\ref{e33}) is strictly positive since $A>B>w>1$. The imaginary modes give rise to two asymptotically degenerate eigenvectors given by
\begin{equation}
\label{e34}
X^{\dag}(iv_{1}) \vert ++  \rangle \quad \mathrm{and} \quad X^{\dag}(iv_{2})  \vert ++ \,\rangle \, ,
\end{equation}
with eigenvalues
\begin{equation}
\label{e35}
\Lambda_{1} = \Lambda_{0} \textrm{e}^{-\gamma(iv_{1})} \quad \mathrm{and} \quad  \Lambda_{2} = \Lambda_{0} \textrm{e}^{-\gamma(iv_{2})} \, .
\end{equation}
Because $\gamma(iv_1)<\gamma(iv_2)<\gamma(0)$ whereas  $\gamma(k)$ corresponding to the real $k$ are all larger then $\gamma(0)$, these two asymptotically degenerate eigenvectors are the lowest excitation states. The asymptotic degeneracy of the transfer matrix spectrum disappears at temperature $\sim  T_w(h_1) - C(h_1)/M$. For $T>T_w(h_1)$ all momenta $k$ are real.

\subsection{Diverging length scale}
\label{sub:lengthscale}
Let us consider the pair-correlation function $\mathcal{G}_M(m,n)=\langle\sigma_{m,l}\sigma_{m,l+n}\rangle - \langle\sigma_{m,l}\rangle\langle\sigma_{m,l+n}\rangle$ of two spins in the $m$th row separated by $n$ columns (see Fig.~\ref{f02}). Using  the transfer matrix  we can write
\begin{equation}
\label{e37}
\mathcal{C}_M(m,n) =  \langle\sigma_{m,l}\sigma_{m,l+n}\rangle = \frac{ {\rm Tr} \left( \textsf{V}^{L-n}  \sigma_{m}^{x} \textsf{V}^{n}  \sigma_{m}^{x} \right) }{ {\rm Tr} \left( \textsf{V}^{L} \right) } \, ,
\end{equation}
where we have imposed periodic boundary conditions in the strip axial direction. For the edge state corresponding to $h_1<0$  and $h_2=-h_1$,  Eq.~(\ref{e37}) reduces in the limit of $L\to \infty$ to
\begin{align}
\label{e38}\nonumber
& \mathcal{C}_{M}(m,n) = \\
&  = \langle ++ \vert X(iv_{1}) \sigma_{m}^{x} (\textsf{V}/\Lambda_{\max})^{n} \sigma_{m}^{x} X^{\dag}(iv_{1}) \vert ++ \rangle \, ,
\end{align}
where $\Lambda_{\max} = \Lambda_{0} \exp[-\gamma(iv_{1})]$. By  applying the spectral decomposition to $\textsf{V}$ we find that the lowest order contributions to $\mathcal{C}_{M}(n)$ come from the one-particle states, i.e., from
\begin{equation}
\label{e39}
\sum_{k \in \Omega_{M}} \Lambda_{k}^{n}   X^{\dag}(k)\left( \vert \Phi_{\infty} \rangle \langle \Phi_{\infty} \vert +  X^{\dag}_{0}\vert \Phi_{\infty} \rangle \langle \Phi_{\infty} \vert X_{0}\right) X(k) \, ,
\end{equation}
where $\Lambda_{k} = \Lambda_{0} \exp[-\gamma(k)]$ is the eigenvalue of $X^{\dag}(k) \vert \Phi_{\infty} \rangle$ as well as of $X^{\dag}(k) X^{\dag}_0 \vert \Phi_{\infty} \rangle$. Therefore we have
\begin{eqnarray}\nonumber
\label{e40}
\mathcal{C}_{M}(m,n) & \simeq &  \sum_{k \in \Omega_{M}} \vert \langle \, ++ \vert X(iv_{1}) \sigma_{m}^{x} X^{\dag}(k) \vert ++ \rangle \vert^{2} \times \\
& \times &\textrm{e}^{-n(\gamma(k)-\gamma(iv_{1}))} \, .
\end{eqnarray}
The spectral sum can be split into a sum over the states with imaginary wave numbers and a remainder stemming from the real wave numbers.
This  generates three different types of contributions. The wave number $k_{1}=iv_{1}$ gives a $n$-independent contribution
\begin{equation}
\label{e41}
{\mathfrak m}_{m}^{2}(M) = \vert \langle \, ++ \vert X(iv_{1}) \sigma_{m}^{x} X^{\dag}(iv_{1}) \vert ++ \, \rangle \vert^{2} \, ,
\end{equation}
which is the formula for the   square of  magnetization  $\mathfrak{m}_{m}(M) = \langle \sigma (m,l) \rangle$ at the row $m$ of the strip~\cite{MS}. From the wave number $k_{2}=iv_{2}$ we have a $n$-dependent contribution
\begin{equation}
\label{e42}
\vert \langle \, ++ \vert X(iv_{1}) \sigma_{m}^{x} X^{\dag}(iv_{2}) \vert ++ \, \rangle \vert^{2} \textrm{e}^{-n(\gamma(iv_{2})-\gamma(iv_{1}))} \, ,
\end{equation}
whereas the real wave numbers generate terms which are proportional to 
\begin{equation}
\label{e43}
\exp[-n(\gamma(k)-\gamma(iv_{1}))] \, .
\vspace{3mm}
\end{equation}
By virtue of the inequality $\gamma(k)>\gamma(iv_{2})>\gamma(iv_{1})$, provided $k$ is real the terms of the form (\ref{e43}) decay on a shorter length scale compared to (\ref{e42}), and therefore they yield subleading corrections beyond the leading decay given [Eq.~(\ref{e42})]. Because in the partial wetting regime (below the wetting temperature)
\begin{equation}
\label{e44}
\min_{k \in \Omega_{M}}\left( \gamma(k) - \gamma(iv_{1}) \right) >  \gamma(iv_{2}) - \gamma(iv_{1}) \, ,
\end{equation}
for $n\to \infty$ we can write 
\begin{equation}
\label{e45}
\mathcal{G}_{M}(m,n) = \mathcal{C}_{M}(m,n) - \mathfrak{m}_{m}^{2}(M) \simeq  \bar{\mathfrak{m}}_{m}^{2}(M) \textrm{e}^{-n/\xi_{\parallel}} \, ,
\end{equation}
where $\simeq$ stands for the omission of subleading terms due to real wave numbers and 
$\bar{\mathfrak{m}}_{m}(M)= \vert\langle ++ \vert X(iv_{1})\sigma_{m}^{x} X^{\dag}(iv_{2})\vert ++ \rangle \vert$.
Let us consider the middle row of the strip. Due to the symmetry $\langle \sigma (m,l) \rangle  = - \langle \sigma (M-m,l) \rangle$, the magnetization in the midpoint $m=M/2$ (for even $M$) is equal to zero. Thus Eq.~(\ref{e45}) implies  that  in the middle of the strip, the spin pair correlations parallel to the edges of the strip decay to zero exponentially. As follows from Eqs.~(\ref{e24}) and (\ref{e32})  (see also Eq.~(\ref{e:4_9})), the  length scale $\xi_{\parallel}$,  on which long-range order is ultimately lost, diverges exponentially fast as $M\to \infty$~\cite{MS}
\begin{equation}
\label{e46}
\xi_{\parallel} = \frac{(Aw-1)(Bw-1)}{2w\sqrt{AB}\left(w-w^{-1}\right)^{2}} w^{M} \, ,
\end{equation}
with  $w=\textrm{e}^{v_0}$. In equation (\ref{e46}) we used the symbol $\simeq$, however I think we can use the equality because we identify $\xi_{\parallel}$ with the leading term for large $M$ in Eq.~(\ref{e36}).  This behavior is similar to the  strips with free boundary conditions we discussed in the context of the action of the distance in Refs.~\cite{AM_2002,AMSV_2017}. However, there is an important difference: the edge-spin pair correlation function calculated  in these references, decays to zero, whereas in  the present case, it decays to a constant $\mathfrak{m}_{1}(M)$ which is different from the edge spontaneous magnetization. Both $\mathfrak{m}_{1}(M)$ and $\bar{\mathfrak{m}}_{1}(M)$ can be calculated (see App.~\ref{app:5}). We find $\mathfrak{m}_{1}(M) = -\mathfrak{m}_e + O(w^{-M}) + O(M^{-3/2}\textrm{e}^{-2M\hat{\gamma}(0)})$, where $\mathfrak{m}_e$ is  the surface magnetization  in the semiinfinite system with a positive surface field 
(result for $\mathfrak{m}_e$ was also obtained using the Pfaffian method~\cite{McCoyWu, McCoyWu_book}):
\begin{equation}
\begin{aligned}
\label{e47}
& \mathfrak{m}_{e} \simeq \frac{w-w^{-1}}{\sqrt{(w-A^{-1})(w-B^{-1})}} +\\
& + \biggl[ \frac{AB}{(A-w)(B-w)} \biggr]^{1/2} \int_{-\pi}^{\pi} \frac{\textrm{d}k}{2\pi} \Bigl[ \cos\delta^{*}(k) - \cos\delta(k) \Bigr] ,
\end{aligned}
\end{equation}
and
\begin{equation}
\label{e48}
\bar{\mathfrak{m}}_{1}(M) = \frac{w-w^{-1}}{\sqrt{(w-A^{-1})(w-B^{-1})}} + O(w^{-M}) \, .
\end{equation}

\subsection{Surface states}
\label{sub:surfst}
\begin{figure}[htbp]
\centering
\includegraphics[width=\columnwidth]{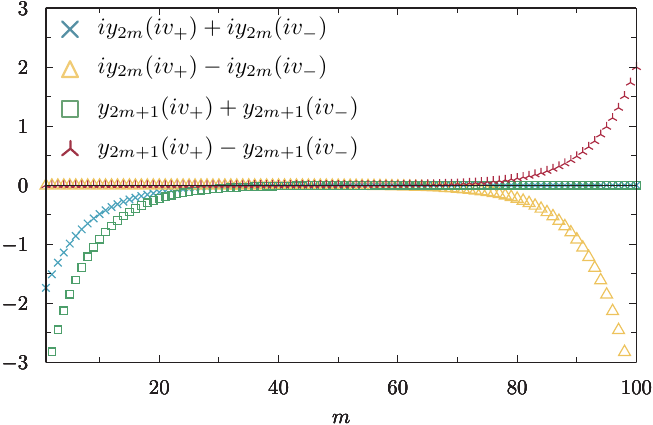}
\caption{Even and odd combinations of eigenvectors of the transfer matrix $\textsf{V}$. In this figure, $K_{1}=K_{2}$, the temperature is $T=2$ and the surface field is $h_{1}=0.3$, corresponding to $w=1.1513$.}
\label{fig_surfacemodes}
\end{figure}

We note that the sums and differences of eigenvectors $\mathcal{E}_{m}^{\pm} = i y_{2m}(iv_1) \pm i y_{2m}(iv_2)$ and $\mathcal{O}_{m}^{\pm} = y_{2m+1}(iv_1) \pm y_{2m+1}(iv_2)$, respectively with $+$ and $-$, decay exponentially from one or the other edge of the strip. In particular, the sums $\mathcal{E}_{m}^{+}$ and $\mathcal{O}_{m}^{+}$ decay exponentially from the bottom edge at $m=0$, whereas the differences  exhibit an exponential decay from the top boundary $m=M+1$; see Fig.~\ref{fig_surfacemodes}.
This can be seen as follows.  By using the discretisation equation, (\ref{e29}) becomes
\begin{equation}
\label{30112023_1210}
y_{2m+1}(iv) = - \textrm{e}^{i\delta^{*}(iv)} \textrm{e}^{-mv} + s \textrm{e}^{-(M-m)v} .
\end{equation}
The above clearly displays a linear combination of exponentially evanescent terms fading out from the two edges. By using (\ref{30112023_1210}), the combination $\mathcal{O}_{m}^{+}$ becomes
\begin{eqnarray} \nonumber
\mathcal{O}_{m}^{+} & = & y_{2m+1}(iv_{1}) + y_{2m+1}(iv_{2}) \\ \nonumber
& = & \biggl[ - \textrm{e}^{-i\delta^{*}(iv_{1})-mv_{1}} - \textrm{e}^{-i\delta^{*}(iv_{2})-mv_{2}} \biggr] \\
& + & \biggl[ \textrm{e}^{-(M-m)v_{1}} - \textrm{e}^{-(M-m)v_{2}} \biggr] \, .
\end{eqnarray}
Since $v_{1}$ and $v_{2}$ are exponentially degenerate for large $M$, the two terms in the first bracket become equal for large large $M$, while the terms in the second square bracket almost cancel each other leaving an exponentially subleading correction. Therefore we obtain $\mathcal{O}_{m}^{+} \simeq -2 \exp(-i\delta^{*}(iv_{0})) \exp(-m v_{0})$, which corresponds to an interface running bound to the edge $m=0$; see the green squares in Fig.~\ref{fig_surfacemodes}. A similar analysis can be carried out for the remaining combinations.

\begin{figure}[htbp]
\centering
\includegraphics[width=\columnwidth]{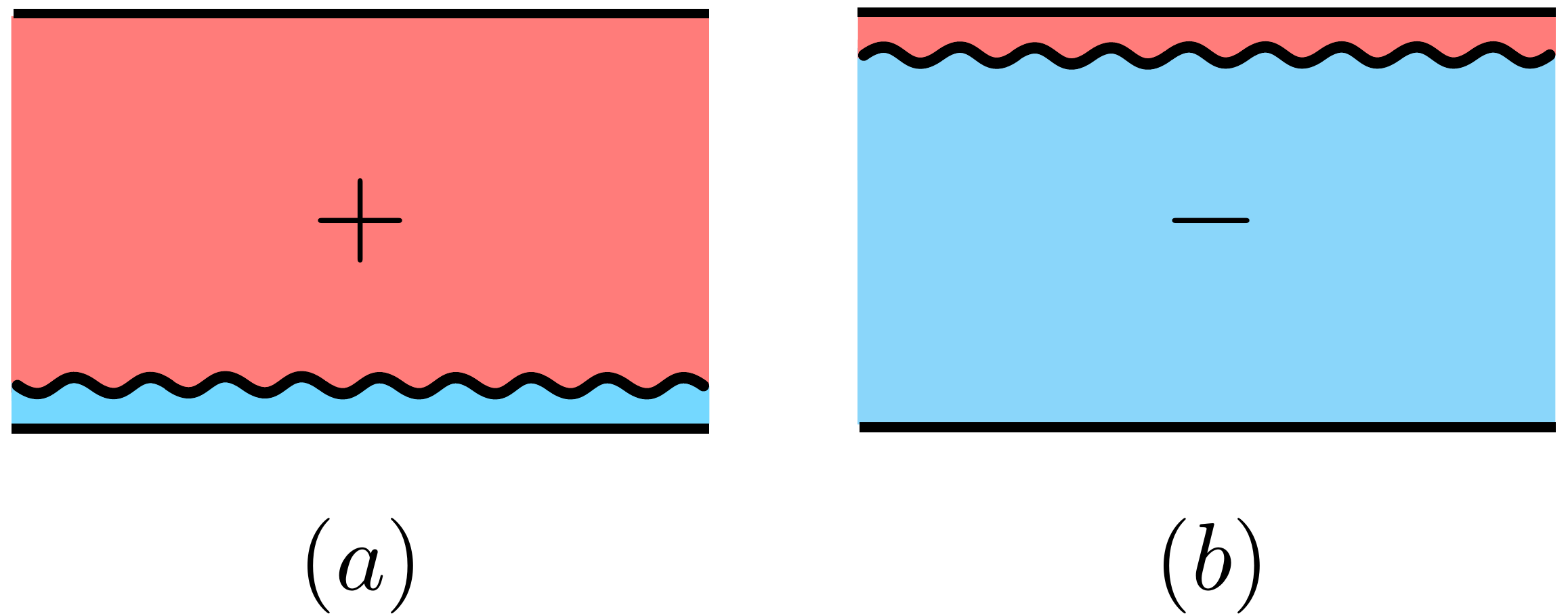}
\caption{Depiction of a domain wall bound to the wall at $m=0$ $(a)$ and at $m=M+1$ $(b)$.}
\label{fig_modes}
\end{figure}

This behavior suggests the construction of the following even/odd combination for putative surface states
\begin{equation}
\begin{aligned}
\label{e49}
\vert e \rangle &= 2^{-1}\Bigl[ X^{\dag}(iv_{1}) + X^{\dag}(iv_{2}) \Bigr](1+ X^{\dag}_0) \vert \Phi_{\infty} \, \rangle\, , \\
\vert o \rangle &= 2^{-1}\Bigl[ X^{\dag}(iv_{1}) - X^{\dag}(iv_{2}) \Bigr](1+ X^{\dag}_0) \vert \Phi_{\infty} \, \rangle\, .
\end{aligned}
\end{equation}
Note that these states are not the eigenstates of the transfer matrix. It can be shown that as $M\to \infty$ they have the property
\begin{equation}
\label{e50}
\langle o \vert \sigma_1^x\vert o \rangle = - \mathfrak{m}_e  \quad \mathrm{and}  \quad \langle e \vert \sigma_M^x \vert e \rangle  =  \mathfrak{m}_e \, .
\end{equation}
We refer to Appendix.~\ref{app:5} for details of the calculations of the edge magnetisations. Thus, $\vert e \rangle$ and $\vert o \rangle$ can be interpreted as states in which the domain wall is bound to the one side of the strip or to the other one, as sketched in Fig.~\ref{fig_modes}.
It is straightforward to show that
\begin{eqnarray}
\label{e51}
\Lambda_{1}^{-n}\langle e \vert \textsf{V}^n\vert e \rangle &=& 2^{-1}\left(1+ \exp(-n/\xi_{\parallel})\right) \, , \\ \nonumber
\Lambda_{1}^{-n}\langle o \vert \textsf{V}^n\vert e \rangle &=& 2^{-1}\left(1- \exp(-n/\xi_{\parallel})\right) \, .
\end{eqnarray}
If $n \gg \xi_{\parallel}$ then the elements of transition matrix tend to $1/2$. This means that the system flips between different surface states on the length scale $\xi_{\parallel}$.

\section{Summary and conclusions}
\label{sec5}

In this paper we have studied a network Ising model constructed from a 2D array of boxes and connecting strips with wetting boundaries. We have showed that in the partial wetting regime, the parameters can be tuned to produce long-range order. If the connecting channels are long enough, then the ordering between boxes extends over many thousands of molecular diameters. We expect the similar scenario for surface fields that are marginally long-ranged~\cite{PhysRevE.75.041110}. The above phenomenon is in sharp contrast with the exponential decay of order in a bulk system in which the decay takes place on the scale of the bulk correlation length. It has been shown rather recently that a classical system supports such a type of order with free boundary conditions \cite{AMSV_2017} and that this effect is not a prerogative of inherently quantum systems. However, the potential feasibility of free boundaries in experiments with binary liquid mixtures at the walls requires fine tuning of the interactions between the walls and the fluid components, because in general the walls tend to be wetted by one component more than the other. This experimental fact motivated the present  study of surface fields acting on the boundaries. One of the most important results we obtained is that the above mentioned long-range ordering known for free boundaries protracts also for surface fields. Our theory, which applies to classical lattice gases and their analogues, maybe tested in experiments and in Monte Carlo simulations~\cite{Oleg}. Our paper also contains new exact results regarding the Ising model in two dimensions. We have given a new microscopic analysis of surface states for the Ising strip with opposing surface fields. Surface states produce an asymptotic degeneracy of the transfer matrix in a partial wetting region~\cite{Kac}. We calculated exactly the free energy associated with a domain wall running at the angle  to the edges of the strip and the point tension for boundaries subject to surface fields.

\section*{Acknowledgements}
A. M. was partially supported by  the Polish National Science Center (Opus Grant  No.~2022/45/B/ST3/00936).
A. S. acknowledges FWF Der Wissenschaftsfonds for funding through the Lise-Meitner Fellowship (Grant No. M 3300-N).
We thank O. A. Vasilyev for sharing with us his preliminary  Monte Carlo simulation results.

\appendix
\section{Contact angle}
\label{app:1}
\begin{figure}[htbp]
\centering
\includegraphics[width=5.7cm]{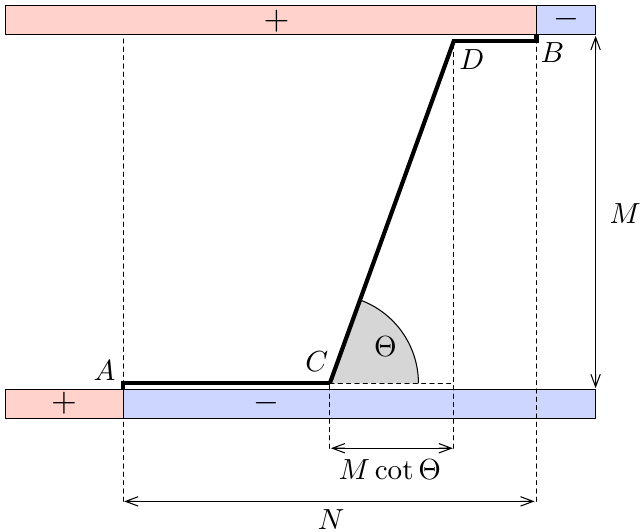}
\caption{A domain wall composed of two  flat pieces running parallel to the wall and  section inclined at an angle $\Theta$ (called wetting angle). The optimal value of $M\cot\Theta$ is obtained by minimizing the excess free energy associated with such an interfacial configuration.}
\label{f07}
\end{figure}
\begin{figure}[htbp]
\centering
\includegraphics[width=0.45\textwidth]{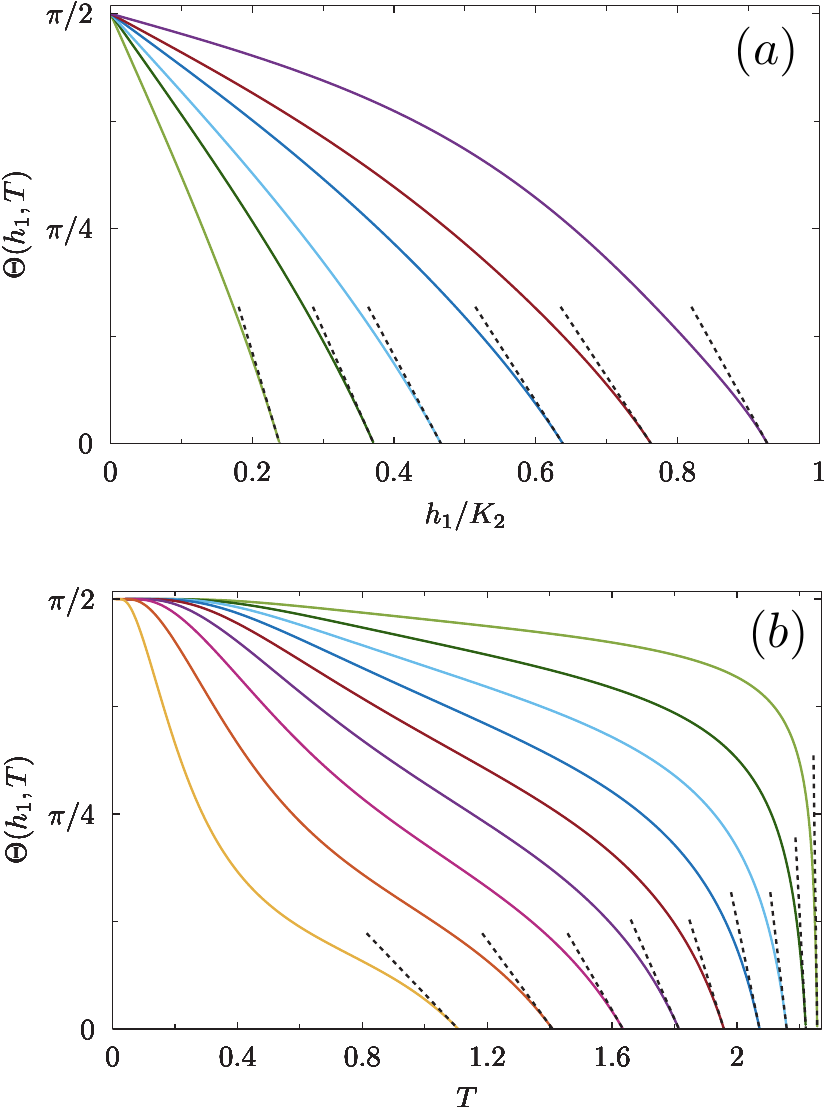}
\caption{The contact angle $\Theta(h_{1},T)$ as a function of the rescaled field $h_{1}/K_{2}$ for fixed temperature $(a)$, and as a function of the temperature for fixed rescaled field $h_{1}/K_{2}=a$. In panel $(a)$ the temperature takes the values $T=2.2$, $2.1$, $2.0$, $1.75$, $1.5$, $1.0$ (from left to right curves). In panel $(b)$ the rescaled field ranges from $0.1$ to $0.9$ with spacing $0.1$ (from top to bottom curves). The dashed black lines in $(a)$ and $(b)$ show the linear vanishing of the contact angle upon approaching the wetting phase boundary. In these figures, $K_{1}=K_{2}$.}
\label{fig02_01}
\end{figure}

Consider a strip with boundary conditions that introduce a domain wall  pinned on two edges, as shown in Fig.~\ref{f07}. The pinning points are offset by  $N$ in horizontal direction and $M$ in vertical direction. In the presence of surface fields at the boundaries, the surface attracts the domain wall, therefore, below the wetting temperature, the domain wall cannot be a straight line connecting the pinning points at the edges of the strip. Rather, it will adopt the shape  shown in Fig.~\ref{f07} with the contact angle that can be determined  from minimization of the excess free energy associated with such a domain wall.
Let $f_{0}$ be the surface free energy of a flat interface pinned  to the wall and let $\tau(\vartheta)$ be the angle-dependent surface tension~\cite{AR74,AU88} for an inclined interface forming a tilt angle $\vartheta$ with the wall as shown  in Fig.~\ref{f07}. The angle-dependent surface tension was calculated in the case where there is no attracting bonds at the edges, i.e., for $\vert h_1 \vert =\vert h_2\vert =K_2$, in Ref.~\cite{AR74}; it satisfies
\begin{equation}
\label{e:1_1}
\tau(\vartheta) = (\cos\vartheta) \gamma(iv_{s}(\vartheta)) + (\sin\vartheta) v_{s}(\vartheta) \, ,
\end{equation}
with $v_{s}(\vartheta)$ given by the saddle point calculations as  ~\cite{AU88}
\begin{equation}
\label{e:1_2}
\gamma^{(1)}(iv_{s}(\vartheta)) =  i\tan\vartheta \, ,
\end{equation}
where here the superscript stands for the first derivative with respect to $k$ of Onsager's $\gamma(k)$ function given by Eq.~(\ref{e24}).
Thus the excess free energy for the domain wall shown in Fig.~\ref{f07} can be written as
\begin{equation}
\label{e:1_3}
\mathcal{F}(\vartheta) = (M \csc\vartheta) \tau(\vartheta) + (N-M\cot\vartheta) f_{0} - N f_{0} \, .
\end{equation}
The first term is due to the inclined interface, the second is due to the horizontal portions. The third term is the free energy for a  flat interface pinned to the wall; the subtraction ensures that $\mathcal{F}(\vartheta)$ is actually the \emph{excess} free energy. Now we look for the wetting angle $\Theta$ which minimizes the function $\mathcal{F}(\vartheta)$. 
By using the identity
\begin{equation}
\label{e:1_4}
\tau^{\prime}(\vartheta) = -(\sin\vartheta) \gamma(iv_{s}(\vartheta)) + (\cos\vartheta) v_{s}(\vartheta) \, ,
\end{equation}
which follows from (\ref{e:1_1}) and (\ref{e:1_2}), we find
\begin{equation}
\label{e:1_5}
\partial_{\vartheta}\mathcal{F}(\vartheta) = M \csc^{2}\vartheta \bigl[f_{0} - \gamma(iv_{s}(\vartheta)) \bigr] \, .
\end{equation}
The stationary condition gives
\begin{equation}
\label{e:1_6}
f_{0} = \gamma(iv_{s}(\Theta)) \, ,
\end{equation}
in agreement with the result of earlier studies with $v_{s}=v_0$~\cite{MS, Abraham_1980}. This,  together with Eq.~(\ref{e:1_1}), lead to a rather compact expression for the excess free energy associated to the insertion of an inclined domain wall
\begin{equation}
\label{e:1_7}
\mathcal{F}(\Theta) = M v_{0}(\Theta) \, .
\end{equation}
Equation (\ref{e:1_2}) gives the contact angle $\Theta$ as the solution of
\begin{equation}
\label{e:1_8}
\frac{s_{1}^{\star}s_{2}\sinh v_{0}(\Theta)}{\sinh\gamma(iv_{0}(\Theta))} = \tan \Theta \, ,
\end{equation}
with the shorthand notation $s_{1} = \sinh 2K_{1}$ and  $s_{2}^{\star}  = \sinh2K_{2}^{\star}$. Hence, if $v_0\searrow 0$ at the wetting transition, then $\Theta \searrow 0$ as anticipated. In Appendix~\ref{app:4} we will demonstrate by exact microscopic calculation of $\mathcal{F}(\Theta)$ that indeed  $v_{s}(\Theta)$ in Eq.~(\ref{e:1_6}) is equal to $v_0=\ln w$.
Figure \ref{fig02_01}  shows the contact angle $\Theta$ as a function of the  surface field and temperature.

\section{Hard rod lattice gas and the Tonks gas}
\label{app:2}
Proceeding with the exact grand partition function of the hard rod lattice gas given by Eq.~(\ref{e3}) is not a simple task. For this reason it makes sense to find a way to simplify the expression of the partition function. At this point it is very useful to show how we can approximate
the hard rod lattice gas with a Tonks-Rayleigh hard rod gas in the continuum~\cite{Tonks,RobledoRowlinson}. Recall that the canonical partition function of  Tonks-Rayleigh hard rod gas with $j$ particles on the line of length $L$ in the absence of any external potential is equal to
\begin{equation}
\label{ }
\mathcal{Q}_j^{\rm Tonks}=\frac{(L-j\sigma+j)^j}{j!} \, ;
\end{equation}
this equation appeared in the seminal paper by Lee and Yang on the theory of equations of state (see Eq.~(52) of \cite{LeeYang1952}). 

By applying Stirling's formula to the binomial $\binom{L-j\sigma+j}{j}$ for fixed $j$ and large $L-j\sigma+j$, we find
\begin{equation}
\label{e:2_1}
\begin{aligned}
\mathcal{Q}_j^{\rm lattice\, HR}& =\binom{L-j\sigma+j}{j} \\
&\approx \frac{\left(L-j\sigma +j\right)^j}{j!}=\mathcal{Q}_j^{\rm Tonks}(L,\sigma-1),
\end{aligned}
\end{equation}
 where $\mathcal{Q}_j^{\rm lattice \, HR}$ is the canonical partition function of the lattice hard rods model. Having established a connection between the canonical partition function for hard rods on the lattice and hards rods in the continuum (Tonks gas), we now show how this enables us to simplify the grand partition function (Eq.~(\ref{e3})). To this end we write the grand partition function of the Tonks gas
\begin{equation}
\label{e:2_2}
\begin{aligned}
&\Xi^{\textrm{Tonks}}(\zeta,L,\sigma) = \sum_{j=0}^{[L/\sigma]} \zeta^{j} (L-j\sigma)^{j}/j! \\ \nonumber
& \approx  \sum_{j=0}^{[L/\sigma]} (\zeta\sigma)^{j} (L/\sigma)^{j}/j!  \approx  \sum_{j=0}^{[L/\sigma]} (\zeta\sigma)^{j} \binom{L\sigma}{j} \\ \nonumber
& =  (1+\zeta\sigma)^{L/\sigma}  =\Xi^{\textrm{lattice gas}}(\zeta\sigma, L/\sigma) \, ,
\end{aligned}
\end{equation}
where $\Xi^{\textrm{lattice gas}}(\zeta, L)=(1+\zeta)^{L}$ is the grand partition function for the hard rod gas with unit diameter encountered in the free edge problem. The approximation then connects the Tonks gas to the lattice gas, thus providing a bridge between the continuum and the discrete descriptions of the one dimensional gases of impenetrable particles. The chain of approximations is valid in the regime of small $\zeta$, which predominantly weights those configurations with small $j$. Therefore at small fugacity the Tonks gas behaves as an ideal lattice gas with enhanced fugacity $\zeta\sigma$ in a reduced volume $L/\sigma$.

\section{Onsager angles}
\label{app:3}
The functions $\delta^{\prime}(k)$ and $\delta^*(k)$, introduced by Onsager \cite{Onsager_44}, are elements of a hyperbolic triangle whose edges have length $K_{1}^{\star}$, $K_{2}$, and $\gamma(k)$. The angle $\delta^*(k)$ is related to the other geometrical elements via
\begin{equation}
\label{ }
c_{1}^{\star} = c_{2} \cosh\gamma(k) - s_{2} \sinh\gamma(k) \cos\delta^{*}(k) \, ,
\end{equation}
which is formally analogous to Eq.~(\ref{e24}). Here, we use the following shorthand notation: $s_{1}^{\star}=\sinh 2K_{1}^{\star}$, $c_{1}^{\star}=\cosh 2K_{1}^{\star}$, $s_{2}=\sinh 2K_{2}$, $c_{2}=\cosh 2K_{2}$. Furthermore, by combining the above with Eq.~(\ref{e24}) we find
\begin{equation}
\label{21112023_1619}
\cos\delta^{*}(k) = \frac{c_{1}^{\star}s_{2}  - s_{1}^{\star}c_{2}\cos k}{\sinh\gamma(k)} \, .
\end{equation}
The above angles admit the factorized expressions
\begin{equation}
\label{e:3_1}
\textrm{e}^{2i\delta^{\prime}(k)} = \frac{\textrm{e}^{ik}-A}{A\textrm{e}^{ik}-1} \frac{\textrm{e}^{ik}-B}{B\textrm{e}^{ik}-1} \, ,
\end{equation}
and
\begin{equation}
\label{e:3_2}
\textrm{e}^{2i\delta^{*}(k)} = \frac{\textrm{e}^{ik}-A}{A\textrm{e}^{ik}-1} \frac{B\textrm{e}^{ik}-1}{\textrm{e}^{ik}-B} \, ,
\end{equation}
Physical arguments demand that $\gamma(0)>0$ for both $T<T_{c}$ and $T>T_{c}$. On the other hand, the behavior at $k=0$ of the Onsager angles depends on the temperature, as can be realized by plugging $k=0$ in Eq.~(\ref{21112023_1619}). For subcritical temperatures ($A>B>1$) the sheet of the square root is selected such that $\delta^{*}(0)=0$, thus $\textrm{e}^{i\delta^{*}(0)}=+1$. For supercritical temperatures ($B<1$) we have $\delta^{*}(0)=\pi$, hence $\textrm{e}^{i\delta^{*}(0)}=-1$. An analogous treatment applies to the angle $\delta^{\prime}(k)$ by noting that it can be obtained by mapping $B$ to $B^{-1}$ in the expression of $\delta^{*}(k)$. Therefore, for subcritical temperatures $\textrm{e}^{i\delta'(0)}=-1$.

\section{Boltzmann weight for opposing surface fields}
\label{app:4}
\begin{figure}[htbp]
\centering
\includegraphics[width=0.45\textwidth]{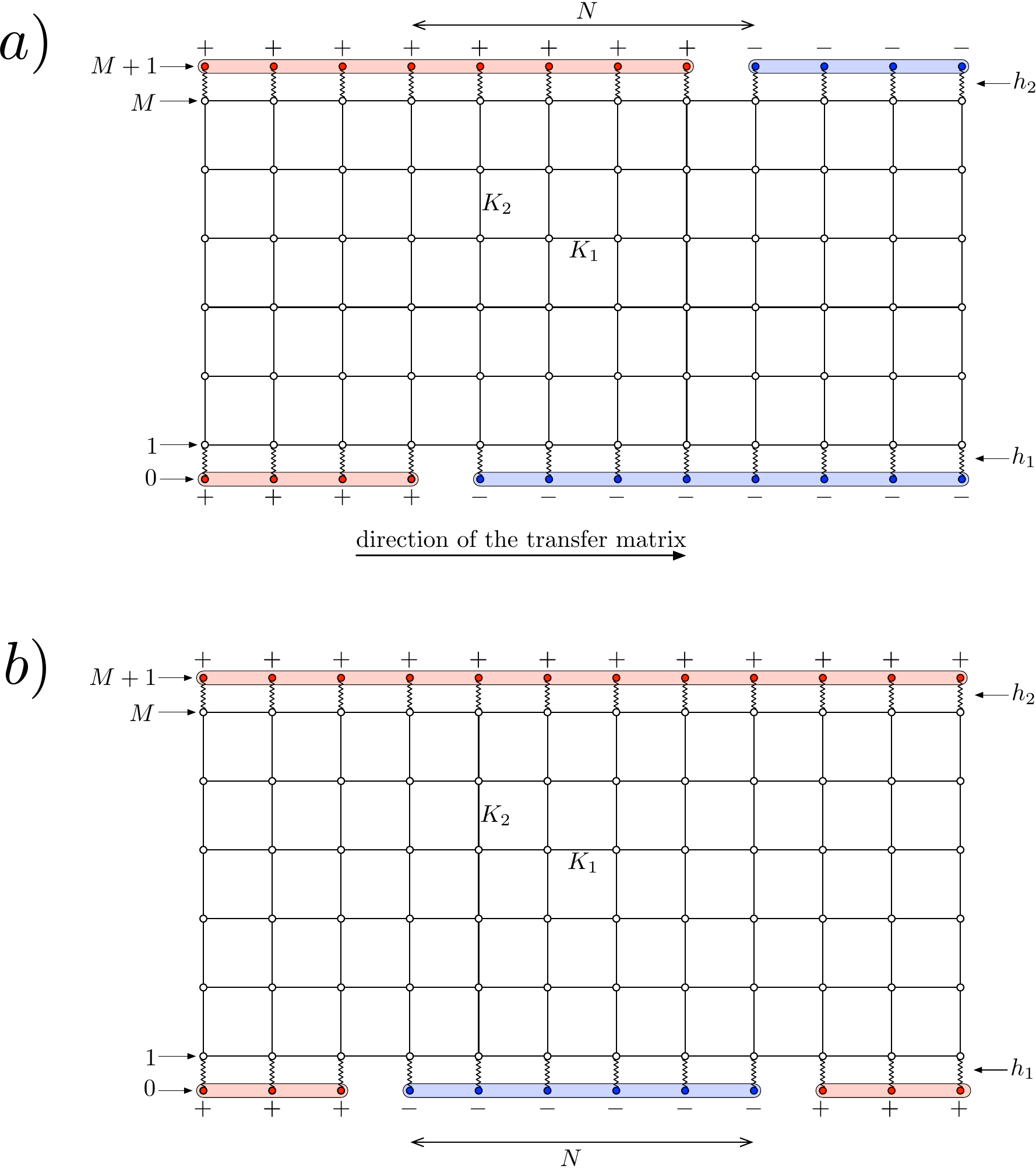}
\caption{Lattice used for the calculation of the free energy for the inclined (upper panel) and ``flat'' (bottom panel) domain walls.}
\label{f08}
\end{figure}

The free energy $F(N,M)$ of the domain wall shown in Fig.~\ref{f07} can be calculated from a canonical partition function ratio $Z^{\times}/Z$ for a system with and without a domain wall (see Fig.~\ref{f08} for clarification of  notations) using a transfer matrix approach
\begin{equation}
 \label{e:4_1}
 Z^{\times}/Z = \frac{ \textrm{Tr} \Bigl[ \left( \textsf{V}^{\prime} \right)^{L-N} \left( -\sigma_{0}^{z} \right) \left( \textsf{V}^{\prime} \right)^{N} \left( -\sigma_{M+1}^{z} \right) \Bigr] }{ \textrm{Tr} \Bigl[ \left( \textsf{V}^{\prime} \right)^{L} \Bigr] } \, ,
\end{equation}
where $L$ stands for the length of the lattice shown in Fig.~\ref{f08}. Technically, the lattice is wrapped onto a cylinder; more correctly, there is another domain wall but it is assumed to be far away.
We have used $\textsf{V}^{\prime}=\textsf{V}_1^{1/2}\textsf{V}_2\textsf{V}_1^{1/2}$ as symmetrisation~\cite{Kaufmann_49}.  This symmetrisation is more convenient here because the rotation operators $-\sigma_{0}^{z}$ and $-\sigma_{M+1}^{z}$ involved in (\ref{e:4_1}) anticommute with $\textsf{V}_{1}$. These operators reverse the spins on the bottom $m=0$ and  top $m=M+1$ edges and thus introduce  a  domain wall across the strip. For the edge state corresponding to $h_1 > 0$ and $h_2 = h_1$, Eq.~(\ref{e:4_1}) reduces in the
limit  $L\to\infty$  to
\begin{equation}
\label{e:4_2}
\textrm{e}^{-F(N,M)} = \frac{ \langle \, ++ \vert \left( - \sigma_{0}^{z} \right) \left( \textsf{V}^{\prime} \right)^{N} \left( - \sigma_{M+1}^{z} \right) \vert -- \, \rangle }{ \langle \, ++ \vert \left( \textsf{V}^{\prime} \right)^{N} \vert ++ \, \rangle } \, ,
\end{equation}
where the in- and out- asymptotic states $\vert ++ \rangle$ and $\vert -- \rangle$ are defined in the same way as in Sec.~\ref{sec:model}, but with $X_0$ in Eq.~(\ref{e27}) replaced by $X_0^{\prime}$ for the $\textsf{V}^{\prime}$ symmetrisation. By using the relations $-\sigma_{0}^{z} = i\sigma_{0}^{x}\sigma_{0}^{y} = i \Gamma_{-1}\Gamma_{0}=(X^{\prime}_0+(X^{\prime}_0)^{\dagger})i\Gamma_0$ and
$-\sigma_{M+1}^{z} = i\sigma_{M+1}^{x}\sigma_{M+1}^{y} = i \Gamma_{2M+1}\Gamma_{2M+2}=\Gamma_{2M+1}(X^{\prime}_0-(X^{\prime}_0)^{\dagger})$ (see Sec.~\ref{sec:model}) and applying the spectral decomposition to the operator $\textsf{V}^{\prime}$, we find that the lowest order  contributions to $\textrm{e}^{-F(N,M)}$ come from the one-particle states, therefore 
\begin{equation}
 \label{e:4_3}
 \begin{aligned}
& \textrm{e}^{-F(N,M)} = \\
& \sum_{k\in \Omega_{M}}\textrm{e}^{-N\gamma(k)} \langle \Phi_{\infty} \vert i\Gamma_{0}(X^{\prime}_k)^{\dagger} \vert \Phi_{\infty}\rangle \langle \Phi_{\infty} \vert X^{\prime}_k\Gamma_{2M+1}\vert \Phi_{\infty} \, \rangle .
\end{aligned}
\end{equation}
By using the expressions for $\Gamma_{0}$ and $\Gamma_{2M+1}$ in terms of Fermi operators (see Eq.~(\ref{e:5_5})), we find
\begin{equation}
 \label{e:4_4}
 \begin{aligned}
& \textrm{e}^{-F(N,M)} =  \\
& =  \sum_{k\in \Omega_{M}}(N^{\prime}(k))^2i \left( y_{0}^{\prime}(k) \right)^{*} y_{2M+1}^{\prime}(k) \textrm{e}^{-N\gamma(k)} \, ,
\end{aligned}
\end{equation}
where the eigenvectors for the $\textsf{V}^{\prime}$ symmetrisation are:
\begin{equation}
\begin{aligned}
\label{e:4_5}
y_{2m-1}^{\prime}(k) = & \textrm{e}^{i\delta^{\prime}(k)}\textrm{e}^{imk} + \textrm{e}^{i\delta(k)}\textrm{e}^{-i(m-1)k}  , \\
iy_{2m}^{\prime}(k) = &  \textrm{e}^{imk} + \textrm{e}^{i\delta(k)}\textrm{e}^{i\delta^{\prime}(k)}\textrm{e}^{-i(m-1)k} ,
\end{aligned}
\end{equation}
for $m=1,\dots,M$, with boundary values  
\begin{equation}
\begin{aligned}
\label{e:4_6}
y_{0}^{\prime}(k) & = & i \frac{\sinh(2h_{1})\cosh K_{1}^{\star}}{\sinh\gamma(k)} y_{1}^{\prime}(k) , \\
y_{2M+1}^{\prime}(k) & = & i \frac{\sinh(2h_{1})\cosh K_{1}^{\star}}{\sinh\gamma(k)} y_{2M}^{\prime}(k) .
\end{aligned}
\end{equation}
The allowed momenta $k$ are the same as for the $\textsf{V}$ symmetrisation and are found as the solutions of Eq.~(\ref{e28}). It is convenient to single out from the spectral sum the contributions from the two imaginary wave numbers and write
\begin{eqnarray}
\label{e:4_7}
\textrm{e}^{-F(N,M)} & = & T(iv_{1}) + T(iv_{2}) + \textrm{(real modes)} \, ,
\end{eqnarray}
where
\begin{equation}
\begin{aligned}
\label{e:4_8}
&T(iv)=(N^{\prime})^2(iv)\left(\frac{\sinh(2h_1)\cosh K_{1}^{\star}}{\sinh\gamma(iv)}\right)^2\textrm{e}^{-N\gamma(iv)}\times \\
&\left[s\textrm{e}^{2i\delta^{\prime}(iv)}\textrm{e}^{-2v}+2\textrm{e}^{i\delta^{\prime}(iv)}\textrm{e}^{-(M+1)v}+s\textrm{e}^{-2Mv}\right].
\end{aligned}
\end{equation}
Since $\gamma(k_1)<\gamma(k_2)<\gamma(k_3)\ldots < \gamma(k_M)$, contributions from the real wavenumber decay faster than those from imaginary modes, and can therefore be neglected in the regime of our interest. In the limit $M\to \infty$ the sum of imaginary terms cancel out because the two imaginary solutions $k_1=iv_1$ and $k_2=iv_2$ are asymptotically degenerate and have the opposite parity number $s$ (defined by Eq.~(\ref{e28})). We expand $T(iv_i)$ around $iv_0$ for large $M$ at fixed $N\gg M$ and keep the leading terms in $M$.
From Eqs.~(\ref{e24}) and (\ref{e32}), we have
\begin{equation}
\label{e:4_9}
\gamma(iv_{i})  \simeq \gamma(iv_{0}) + s/(2\xi_{\parallel})   \quad (i=1,2) \, ,
\end{equation}
and thus
\begin{equation}
\label{e:4_10}
\textrm{e}^{-N\gamma(iv_{i})}  \simeq \textrm{e}^{-N\gamma(iv_{0})} \left(1+Ns/(2\xi_{\parallel})  \right)  \quad (i=1,2) \, ,
\end{equation}
where $s=+1$ for $v_1$ and $s=-1$ for $v_2$ and $\xi_{\parallel}$ is given by Eq.~(\ref{e46}). In the limit $M\to \infty$, the normalization constant takes the following form
\begin{equation}
\begin{aligned}
\label{e:4_11}
& (N^{\prime})^{2}(iv)\simeq \frac{w^2-1}{2\textrm{e}^{2i\delta^{\prime}(iv_0)}} \\
&- 2sw^{-1}\textrm{e}^{i\delta^{\prime}(iv_0)}\left(\frac{w^2-1}{2\textrm{e}^{2i\delta^{\prime}(iv_0)}}\right)^2M\textrm{e}^{-Mv_0} \, ,
\end{aligned}
\end{equation}
and the prefactor multiplying the square brackets in Eq.~(\ref{e:4_8}) can be factorized:
\begin{equation}
\label{e:4_12}
\frac{\sinh(2h_1)\cosh K_{1}^{\star}}{\sinh\gamma(iv_0)}=w\sqrt{\frac{AB-1}{(Aw-1)(Bw-1)}} \, .
\end{equation}
Neglecting all terms of the order of $\textrm{e}^{-Mv_0}$ and  higher as subdominant with respect to $M\textrm{e}^{-Mv_0}$ and using Eq.~(\ref{e:3_1}), we find
\begin{equation}
 \begin{aligned}
\label{e:4_13}
& T(iv_1)+T(iv_2) \simeq \\
&-\textrm{e}^{-N\gamma(iv_{0})-Mv_0} \frac{\sqrt{AB}(AB-1)w^2(w-w^{-1})^3}{(Aw-1)^2(Bw-1)^2} \times \\ 
& \biggl[ N-M \frac{\left[(Aw-1)(Bw-1)(w-A^{-1})(w-B^{-1})\right]^{1/2}}{w^2-1} \biggr] \, .
\end{aligned}
\end{equation}
The contact angle $\Theta$ can be identified by expressing Eq.~(\ref{e:1_8}), which is an implicit equation for $\Theta$, in the factorised form
\begin{equation}
\begin{aligned}
 \label{e:4_14}
\frac{\sqrt{(w-A)(w-B)(w-A^{-1})(w-B^{-1})}}{w^2-1}=\cot \Theta \, .
\end{aligned}
\end{equation}
The final result for the free energy $F(N,M)$ of the  domain wall can be written as 
\begin{eqnarray}\nonumber
\label{e:4_15}
\textrm{e}^{-F(N,M)} & = & (N-M\cot\Theta)\underbrace{\frac{(AB-1)(w^2-1)}{(Aw-1)(Bw-1)} \textrm{e}^{-N\gamma(iv_{0})}}_{ \textrm{horizontal} } \nonumber \\
&\times &\textrm{e}^{-2\tau_{p}} \textrm{e}^{-Mv_{0}} \, , 
\end{eqnarray}
where
\begin{equation}
\begin{aligned}
 \label{e:4_16}
&\textrm{e}^{-2\tau_{p}}=\frac{w\sqrt{AB}(w^2-1)^2}{(Aw-1)(Bw-1)} \, .
\end{aligned}
\end{equation}

\begin{figure}[htbp]
\centering
\includegraphics[width=0.45\textwidth]{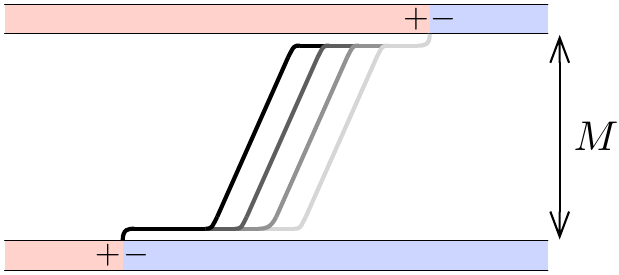}
\caption{Displacements of the inclined portion of the domain wall which do not alter the free energy.}
\label{f09}
\end{figure}

Let us analyze the meaning of the various factors appearing in Eq.~(\ref{e:4_15}). The quantity in parentheses is the entropic factor, which gives the number of ways in which it is possible to displace the inclined portion of the domain wall without altering the free energy cost -- see  Fig.~\ref{f09} for an illustration.

Additional calculations show  that the contribution labeled ``horizontal'' is  due to a flat portion of a domain wall pinned  to the surface with fixed ends (lines AC and DB in in Fig.~\ref{f07}) with fixed ends. Since $N\gamma(iv_{0})=Nf_0$ (see Eq.~(\ref{e:1_6})) is the free energy of such domain wall with length $N$, the other factor must be the contribution from the end points  (points A and B in in Fig.~\ref{f07}). For additional calculations we have considered a lattice with spins reversed between $1$ and $N$ at a bottom edge of the strip as shown in Fig.~\ref{f08}(b)  to introduce  a domain wall running parallel to the  (1,0) axis. The free energy of such domain wall is given by 
Eq.~(\ref{e:4_1}) but with   $-\sigma^z_{M+1}$  replaced by $-\sigma^z_{0}$, and  can be computed in the similar way as above.
The final result reduces in the limit of $ M \to \infty$ at fixed and large $N$  to the expression labeled as ``horizontal'' in Eq.~(\ref{e:4_15}).
The interpretation of the remaining terms is now straightforward. $Mv_0$ is the excess free energy corresponding to the cost of
 replacing  a piece of  flat domain wall of length  $M\cot\Theta$ (line CD in  Fig.~\ref{f07}) by the  inclined one. Thus we can write
\begin{eqnarray} \nonumber
\label{e:4_17}
Mv_{0}  =   M\csc\Theta \, \tau(\Theta)-M\cot\Theta f_{0}   \, , \\
\end{eqnarray}
where $\tau(\Theta)$ is the angle-dependent surface tension at the wetting angle. This agrees with expression (\ref{e:1_3}) for $\mathcal{F}(\vartheta)$ from the free energy considerations in Appendix~\ref{app:1}. Finally, $\tau_{p}$ in Eq.~(\ref{e:4_16}) can be interpreted as a point tension (contributions from points C and D in Fig.~\ref{f07}).

\section{The surface states and edge magnetization}
\label{app:5}
Here we outline calculations of $\langle o \vert \sigma_1^x \vert o \rangle$ and $\langle o \vert \sigma_M^x \vert o \rangle$ and of the edge magnetization for equal and opposing surface fields. First, we express the spin operators in terms of spinors. From Eqs~(\ref{e19}) and  (\ref{e20}) we find $\sigma_1^x=i\Gamma_{-1}\Gamma_0\Gamma_1$ with $\Gamma_{-1}= X_0+X^{\dag}_0$ and $\sigma^x_M=-i\Gamma_{2M}\Gamma_{2M+1}\mathcal{P}_{M+1}(X_0-X_0^{\dagger})$. $\mathcal{P}_{M+1}$ is the parity operator given by Eq.~(\ref{e20}) with $m=M+1$. We need to know the action of $\mathcal{P}_{M+1}$ on the vacuum and the excited states.
It can be showed that $\mathcal{P}_{M+1}\vert \Phi_{\infty}\rangle = - \vert \Phi_{\infty}\rangle $ and
$\mathcal{P}_{M} X^{\dag}(k_{1}) \dots X^{\dag}(k_{n}) \vert\Phi_{\infty}\rangle = (-1)^{n+1} X^{\dag}(k_{1}) \dots X^{\dag}(k_{n}) \vert\Phi_{\infty}\rangle$.
Using definitions  of the surfaces states (Eq.~(\ref{e49})), we find
\begin{equation}
\label{e:5_1}
\begin{aligned}
 \langle o \vert \sigma_1^x \vert o \rangle & = -\frac{1}{2}\left(A_1+A_2-B_{12}-B_{21}\right), \\
 \langle e \vert \sigma_M^x \vert e \rangle & = \frac{1}{2}\left(\tilde{A}_1+\tilde{A}_2+\tilde{B}_1+\tilde{B}_2\right) \, ,
 \end{aligned}
\end{equation}
where for $i=1,2$
\begin{equation}
\label{e:5_2} 
\begin{aligned}
&A_i  = \langle \Phi_{\infty} \vert X(iv_{i})i\Gamma_0 \Gamma_1 X^{\dag}(iv_{i}) \vert \Phi_{\infty} \, \rangle  \\
&\tilde{A}_i  = \langle \Phi_{\infty} \vert X(iv_{i})i\Gamma_{2M} \Gamma_{2M+1} X^{\dag}(iv_{i}) \vert \Phi_{\infty} \, \rangle \, ,
\end{aligned}
\end{equation}
and for $i,j=1,2$ and $i\ne j$
\begin{equation}
\label{e:5_3} 
\begin{aligned}
&B_{ij}  = \langle \Phi_{\infty} \vert X(iv_{i})i\Gamma_0 \Gamma_1 X^{\dag}(iv_{j}) \vert \Phi_{\infty} \, \rangle  \quad  \\
&\tilde{B}_{ij}  = \langle \Phi_{\infty} \vert X(iv_{i})i\Gamma_{2M} \Gamma_{2M+1} X^{\dag}(iv_{j}) \vert \Phi_{\infty} \, \rangle \, .
\end{aligned}
\end{equation}
In order to evaluate these form factors, we employ the relation between spinors and fermionic operators
\begin{equation}
\label{e:5_4}
\Gamma_m=\sum_{k\in \Omega_{M}}N(k)\left(y_m(k)(X_k)^{\dagger}+(y_m)^*(k)X_k\right) \, ,
\end{equation}
which corresponds to the inversion of Eq.~(\ref{e22}). This gives
\begin{equation}
\label{e:5_5} 
\begin{aligned}
A_{i} &= -i N^2(iv_i) y_1(iv_i)y_0^*(iv_i) + i \sum_{k \ne i} N^2(k) y_1(k)y_0^*(k)  \\
\tilde{A}_i& = -i N^2(iv_i) y_1(iv_i)y_0^*(iv_i) - i \sum_{k \ne i} N^2(k) y_1^*(k)y_0(k) \, ,
\end{aligned}
\end{equation}
for $i,j=1,2$ and 
\begin{equation}
\label{e:5_6} 
\begin{aligned}
B_{ij} & = i N(iv_i)N(iv_j) \left[y_0(iv_i)y_1^*(iv_j) - y_0^*(iv_j)y_1(iv_i)\right] \\
\tilde{B}_{ij}& = i N(iv_i)N(iv_j)\left[ y_1(iv_i)y_0^*(iv_j) -  y_1^*(iv_j)y_0(iv_i)\right] \, ,
\end{aligned}
\end{equation}
for $i,j=1,2$ and $i\ne j$, where we have used the reflection symmetry (Eq.~(\ref{e31})) to express the eigenvectors $y_{2M}$ and $y_{2M+1}$
in terms of  $y_0(k)$ and $y_1(k)$; the latter ones are  given in Eq.~(\ref{e30}).

Now we take the limit  $M\to \infty$ in which  $v_1, v_2 \to v_0$. In this limit the contributions from imaginary wave numbers cancel each other in $A_i$ and $\tilde{A_{i}}$
 and we have 
 \begin{equation}
\label{e:5_7}
\begin{aligned}
 & A_1 = A_2  = \tilde{A_{1}} = \tilde{A_{2}}  = 2\sqrt{\frac{AB}{(A-w)(B-w)}} \\
 & \times\lim_{M\to\infty}\sum_{k \in \rm real} N^2(k)\left(\cos\delta^*(k)-\cos\delta(k)\right) \, .
\end{aligned}
\end{equation}
In order to calculate $B_{ij}$ and $\tilde{B}_{ij}$, we use the quantization condition [Eq.~(\ref{e28})] and eliminate  $\textrm{e}^{i\delta}$ from  the eigenvectors $y_0(iv_i)$ and $y_1(iv_i)$.
Then we take the limit of $M\to \infty$ to find 
 \begin{equation}
 \label{e:5_8}
 \begin{aligned}
 &B_{ij}=-\tilde{B}_{ij} = -2N^2(iv_0)\sqrt{\frac{AB}{(A-w)(B-w)}} \textrm{e}^{-i\delta^{\star}(iv_0)} \\
 & =\frac{w-w^{-1}}{\sqrt{(A^{-1}-w)(B^{-1}-w)}} \, .
 \end{aligned}
\end{equation}
The second line in the above equation is obtained using the asymptotic form of the normalization constant:
\begin{equation}
\label{e:5_9}
\begin{aligned}
&N^{-2}(iv_0)e^{i\delta^{\star}(iv_0)}=\frac{B}{B-w}\textrm{e}^{i\delta^{*}(iv_0)}+\frac{A}{A-w}\textrm{e}^{-i\delta^{*}(iv_0)} \\
&+ \frac{1}{w^2-1}\left(\textrm{e}^{i\delta^{*}(iv_0)} + \textrm{e}^{-i\delta^{*}(iv_0)}\right) \, .
\end{aligned}
\end{equation}
Bringing together all contributions and taking the limit $M\to \infty$ of the sum over the real wavenumbers\footnote{Performing limit $M\to\infty$
of the sum over the real wave numbers $k$ is not straightforward, because they are uniformly distributed between 0 and $\pi$.
},  we finally obtain:
\begin{equation}
\label{e:5_10}
\begin{aligned}
&\langle o \vert \sigma_1^x \vert o \rangle  = \frac{w-w^{-1}}{\sqrt{(A^{-1}-w)(B^{-1}-w)}} -\sqrt{\frac{AB}{(A-w)(B-w)}} \\
&\times \int_{-\pi}^{\pi} \frac{\textrm{d}k}{2\pi} \Bigl[ \cos\delta^{*}(k) - \cos\delta(k) \Bigr] \, ,
\end{aligned}
\end{equation}
and
\begin{equation}
\label{e:5_11}
\begin{aligned}
&\langle e \vert \sigma_M^x \vert e \rangle = -\frac{w-w^{-1}}{\sqrt{(A^{-1}-w)(B^{-1}-w)}}+\frac{\sqrt{AB}}{(A-w)(B-w)}  \\
&\times \int_{-\pi}^{\pi} \frac{\textrm{d}k}{2\pi} \Bigl[ \cos\delta^{*}(k) - \cos\delta(k) \Bigr] \, .
\end{aligned}
\end{equation}

In order to demonstrate the relations (\ref{e51}), let us calculate the  edge magnetization $\mathfrak{m}_{e}(\alpha,\beta)$, where $\alpha = \pm 1$ is the sign of spins  at the bottom edge and  $\beta=\pm 1$ is the sign of  spins fixed by the top edge - they are fixed by the surface field $h_1$ and $h_2$, respectively. For the bottom edge we have
\begin{equation}
\label{e:5_12}
\mathfrak{m}^b_e(\alpha,\beta)= \frac{\textrm{Tr}\left(\textsf{V}^LP_0(\alpha)\sigma_1^xP_{M+1}(\beta)\right)}{\textrm{Tr}\left(\textsf{V}^LP_0(\alpha)P_{M+1}(\beta)\right)},
\end{equation}
where the projection operators are:
\begin{equation}
\label{e:5_13}
\begin{aligned}
P_{0}(\alpha) & = \frac{\alpha}{2}\left( I+\alpha (X_0+X_0^{\dagger})\right) \\
P_{M+1}(\beta)&  = \frac{1}{2}\left( I+\beta (X_0^{\dagger}-X_0)\mathcal{P}_{M+1}\right) \, .
\end{aligned}
\end{equation}
Proceeding just like in the calculations for the surface states, we find
\begin{equation}
\label{e:5_14}
\begin{aligned}
&\mathfrak{m}^b_e(+,-)=-\mathfrak{m}^b_e(-,+)=\langle \phi_{\infty}|X(iv_1)i\Gamma_0\Gamma_1X^{\dagger}(iv_1)|\phi_{\infty}\rangle  \\
&=\sqrt{\frac{AB}{(A-w)(B-w)}} \int_{-\pi}^{\pi} \frac{\textrm{d}k}{2\pi}\left(\cos\delta^*(k)-\cos\delta(k)\right) \, ,
 \end{aligned}
\end{equation}
and
\begin{equation}
\label{e:5_15}
\begin{aligned}
&\mathfrak{m}^b_e(+,+)=-\mathfrak{m}^b_e(-,-)=\langle \phi_{\infty}|i\Gamma_0\Gamma_1|\phi_{\infty}\rangle = \\
&-\frac{w-w^{-1}}{\sqrt{(A^{-1}-w)(B^{-1}-w)}} \\
&+\sqrt{\frac{AB}{(A-w)(B-w)}}\int_{-\pi}^{\pi} \frac{\textrm{d}k}{2\pi}\left(\cos\delta^*(k)-\cos\delta(k)\right) \, .
 \end{aligned}
\end{equation}
For the top edge we have
\begin{equation}
\label{e:5_16}
\mathfrak{m}^t_e(\alpha,\beta)= \frac{\textrm{Tr}\left(\textsf{V}^LP_0(\alpha)\sigma_M^xP_{M+1}(\beta)\right)}{\textrm{Tr}\left(\textsf{V}^LP_0(\alpha)P_{M+1}(\beta)\right)}.
\end{equation}
An analogous evaluation gives
\begin{equation}
\label{e:5_17}
\begin{aligned}
&\mathfrak{m}^t_e(-,+)=-\mathfrak{m}^t_e(+,-) \\ 
&=\langle \phi_{\infty}|X(iv_1)i\Gamma_{2M}\Gamma_{2M+1}X^{\dagger}(iv_1)|\phi_{\infty}\rangle  \\
& =\frac{\sqrt{AB}}{(A-w)(B-w)}\int_{-\pi}^{\pi} \frac{\textrm{d}k}{2\pi} \left(\cos\delta^*(k)-\cos\delta(k)\right) \, .
 \end{aligned}
\end{equation}
We can see that $\mathfrak{m}^t_e(-,+)=-\mathfrak{m}^b_e(-,+)$. On the other hand 
\begin{equation}
 \label{e:5_18}
 \begin{aligned}
&\mathfrak{m}^t_e(+,+)=-\mathfrak{m}^t_e(-,-)=\langle \phi_{\infty}|i\Gamma_{2M}\Gamma_{2M+1}|\phi_{\infty}\rangle = \\
&-\frac{w-w^{-1}}{\sqrt{(A^{-1}-w)(B^{-1}-w)}} \\
&+\frac{\sqrt{AB}}{(A-w)(B-w)} \int_{-\pi}^{\pi} \frac{\textrm{d}k}{2\pi}\left(\cos\delta^*(k)-\cos\delta(k)\right),
 \end{aligned}
\end{equation}
thus
\begin{equation}
\label{ }
\mathfrak{m}^t_e(\pm,\pm)=\mathfrak{m}^b_e(\pm,\pm)=\mathfrak{m}_e
\end{equation}
Comparing results for the edge magnetization with expressions for the surface states expectation values Eqs.~(\ref{e:5_10}) and (\ref{e:5_11}), we arrive at the relation  (\ref{e51}).

\section{Solution of the discretisation equation}
\label{app:F}
\begin{figure}[htbp]
\centering
\includegraphics[width=\columnwidth]{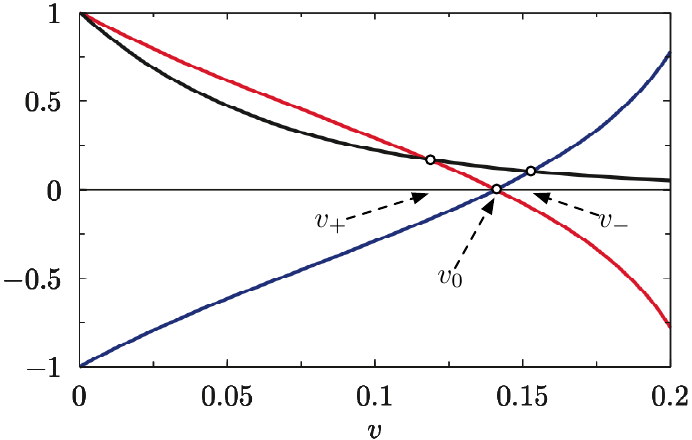}
\caption{Discretisation equation for imaginary wave numbers. The left hand side, $\exp(-Mv)$, is indicated with the solid black curve. The right hand side, $s \exp(i \delta(iv))$, is shown with a solid red ($s=+1$) and blue line ($s=-1$). The corresponding solutions $v_{+}$ and $v_{-}$ are shown. In this figure, $M=15$ and the other parameters are the same as in Fig.~\ref{fig_surfacemodes}.}
\label{fig_disc}
\end{figure}

Here we show how the discretisation equation [Eq.~(\ref{e28})] admits two nearly degenerate imaginary solutions- see Ref.~\cite{MS} where it was originally found. In order to proceed we set $k=iv$ with positive $v$. The left hand side is the exponential $\exp(-Mv)$. The right hand side, $s\exp(i \delta(iv))$ requires special care because of the branch cut exhibited by the Onsager angle $\exp(i \delta^{\prime}(k))$. Since we are interested in the regime $T<T_{c}$, the branch is selected such that $\exp(i \delta^{\prime}(0))=-1$ because $\delta^{\prime}(0)=\pi$. Therefore, the right hand side of the discretisation equation for a wave number along the imaginary axis is
\begin{equation}
\label{24112023_1206}
\textrm{e}^{i \delta(iv)} = - \left( \frac{A \textrm{e}^{v}-1}{A-\textrm{e}^{v}} \frac{B \textrm{e}^{v}-1}{B-\textrm{e}^{v}} \right)^{1/2} \frac{\textrm{e}^{v}-w}{w\textrm{e}^{v}-1} \, .
\end{equation}
Since we are interested in large values of $M$, the left hand side is exponentially small; hence, the solution has to be found in the closeness of the zero of the right hand side. The zero occurs when the second factor of (\ref{24112023_1206}) vanishes, which is at $v=\ln w \equiv v_{0}$. In view of the large-$M$ asymptotic result we need, it is sufficient to perform a Taylor expansion of (\ref{24112023_1206}) around $v=v_{0}$ of the second factor in (\ref{24112023_1206}), which reads
\begin{equation}
\label{ }
\frac{\textrm{e}^{v}-w}{w\textrm{e}^{v}-1} = \frac{v-v_{0}}{w-w^{-1}} + O((v-v_{0})^{2}) \, , \qquad v \rightarrow v_{0} \, .
\end{equation}
It is thus clear that the equation we need to solve is of the form
\begin{equation}
\label{24112023_1218}
\exp(-Mv) = s Q (v-v_{0}) + O((v-v_{0})^{2}) \, ,
\end{equation}
where
\begin{equation}
\label{24112023_1214}
Q = - \left( \frac{A w-1}{A-w} \frac{B w-1}{B-w} \right)^{1/2} \frac{1}{w-w^{-1}}
\end{equation}
can be identified as the prefactor of $v-v_{0}$ appearing in (\ref{24112023_1206}) evaluated at $v=v_{0}$. In Fig.~\ref{fig_disc} we show the left hand side of the discretisation equation together with the right hand hand side with $s=+1$ and $s=-1$.
The inclusion of additional terms beyond the linear one is necessary in order to work out an iterative solution beyond the leading order. By focusing on the leading order term, the solution of (\ref{24112023_1218}) is
\begin{equation}
\label{ }
v = v_{0} + s Q^{-1} \textrm{e}^{-Mv_{0}} + O(\textrm{e}^{-2Mv_{0}}) \, .
\end{equation}
It turns out that $Q$ defined above is related to the quantity $\mathcal{A}$ defined in Eq.~(\ref{e33}) via $Q = - \mathcal{A}^{-1}$. Neglecting the exponentially subleading terms of order $\exp(-2Mv_{0})$, the solution is
\begin{equation}
\label{ }
v_{s} = v_{0} - s \mathcal{A} \textrm{e}^{-Mv_{0}} \, ,
\end{equation}
with $s=+1$ and $s=-1$. The solution closest to the real axis is the one with parity number $s=+1$.

\bibliographystyle{unsrt}
\bibliographystyle{apsrev4-1}
\bibliography{bibliography}

\begin{thebibliography}{39}
\bibitem [{Ole()}]{Oleg}%
  \BibitemOpen
  \href@noop {} {}\bibinfo {note} {Preliminary MC simulation results obtained
  by O.~A.~Vasilyev for a magnetic susceptibility of a 2D array of squares
  connected by strips with opposing surface fields $h_{1}=-h_{2}$ show features
  signaling network ordering below the wetting temperature $T_{w}(h_{1})$,
  which are consistent with our theory.}\BibitemShut {Stop}%
\end{thebibliography}
\widetext
\end{document}